\documentclass[journal]{IEEEtran}

\IEEEoverridecommandlockouts
\usepackage{cite}
\usepackage{amsmath,amssymb,amsfonts}
\usepackage{multirow}
\usepackage{algorithm}
\usepackage{algorithmic}
\usepackage{graphicx}
\usepackage{textcomp}
\usepackage{xcolor}
\usepackage[font=small,labelfont=bf]{caption}
\usepackage{array, makecell}
\newcommand\fullwidth{2.0} % use 2.0 for double column and 1.0 for single column 
\newcommand\halfwidth{1.0} % use 1.0 for double column and 0.5 for single column 

\begin{document}

\title{A Deep Ensemble-based Wireless Receiver Architecture for Mitigating Adversarial Attacks in Automatic Modulation Classification}
% \title{A Deep Ensemble-based Wireless Receiver Architecture for Mitigating Adversarial Interference in Automatic Modulation Classification}
% \title{Assorted Deep Ensembles for Mitigating Automatic Modulation Classification Adversarial Attacks}
%\title{Defending Black Box Adversarial Interference in Deep Learning Signal Classification Receivers}

\author{Rajeev~Sahay,~\IEEEmembership{Student Member,~IEEE,} 
        Christopher~G.~Brinton,\\~\IEEEmembership{Senior~Member,~IEEE,}
        and~David~J.~Love,~\IEEEmembership{Fellow,~IEEE}% <-this % stops a space
\thanks{R. Sahay, C.G. Brinton, and D. J. Love are with the School
of Electrical and Computer Engineering, Purdue University, West Lafayette,
IN, 47907 USA. E-mail: \{sahayr,cgb,djlove\}@purdue.edu.}% <-this % stops a space
\thanks{This work was supported in part by the Naval Surface Warfare Center Crane Division, in part by the Office of Naval Research (ONR) under grant N00014-21-1-2472, and in part by the National Science Foundation (NSF) under grants CNS1642982, CCF1816013, and AST2037864.}%
\thanks{A preliminary version of this material appeared in the Proceedings of the 2021 IEEE International Conference on Communications (ICC) \cite{icc_frq}.}%
}


\maketitle
\begin{abstract}

Deep learning-based automatic modulation classification (AMC) models are susceptible to adversarial attacks. Such attacks inject specifically crafted wireless interference into transmitted signals to induce erroneous classification predictions. Furthermore, adversarial interference is transferable in black box environments, allowing an adversary to attack multiple deep learning models with a single perturbation crafted for a particular classification model. In this work, we propose a novel wireless receiver architecture to mitigate the effects of adversarial interference in various black box attack environments. We begin by evaluating the architecture uncertainty environment, where we show that adversarial attacks crafted to fool specific AMC DL architectures are not directly transferable to different DL architectures. Next, we consider the domain uncertainty environment, where we show that adversarial attacks crafted on time domain and frequency domain features to not directly transfer to the altering domain. Using these insights, we develop our Assorted Deep Ensemble (ADE) defense, which is an ensemble of deep learning architectures trained on time and frequency domain representations of received signals. Through evaluation on two wireless signal datasets under different sources of uncertainty, we demonstrate that our ADE obtains substantial improvements in AMC classification performance compared with baseline defenses across different adversarial attacks and potencies.

%fool models trained on domain specific features are not easily transferable to models trained using features selected from alternate domains. Furthermore, we show that, among models trained using features selected from the same domain, adversarial interference is not directly transferable to AMC classifiers that differ in network architecture compared to the classifier used to craft the attack. %In this capacity, we demonstrate classification performance improvements greater than 30\% on recurrent neural networks (RNNs) and greater than 50\% on convolutional neural networks (CNNs). We further demonstrate our frequency feature-based classification models to achieve accuracies greater than 99\% in the absence of attacks.
% we show that our defense mitigates the effects of transferable black box adversarial attacks in deep learning AMC receivers 
\end{abstract}


\section{Introduction}

\IEEEPARstart{T}{he} recent exponential growth of wireless traffic has resulted in a crowded radio spectrum, which, among other factors, has contributed to reduced mobile efficiency. With the number of devices requiring wireless resources projected to continue increasing, this inefficiency is expected to present large-scale challenges in wireless communications. Automatic modulation classification (AMC), which is a part of cognitive radio technologies, aims to alleviate the inefficiency induced in shared spectrum environments by dynamically extracting meaningful information from massive streams of wireless data. Traditional AMC methods are based on maximum-likelihood approaches \cite{ml1,ml2,ml3,ml4,ml5}, which consist of deriving statistical decision boundaries using hand-crafted (i.e., manually-defined) features to discern various modulation constellations. More recently, deep learning (DL) has become a popular alternative to maximum-likelihood methods for AMC, since it does not require manual feature engineering to attain high classification performance \cite{amc_dl1,amc_dl2,amc_dl5,amc_rnn,tccn_vis}. %\cite{amc_dl1}. 
%  \cite{ml1,ml2,ml3,ml4,ml5,ml_review}

Despite their ability to obtain strong AMC performance, however, deep learning models are highly susceptible to adversarial evasion attacks \cite{adv_filters,amc_adv_atk1,amc_adv_atk2,amc_adv_atk4,amc_adv_atk5}, which introduce additive wireless interference into radio frequency (RF) signals to induce erroneous behavior on well-trained spectrum sensing models. Adversarial interference signals are specifically crafted to alter the classification decisions of trained DL models using a minimum, and often undetectable, amount of power. Not only do adversarial interference signals inhibit privacy and security in cognitive radios, but they are also more efficient than traditional jamming attacks applied in communication networks \cite{jamming}. As a result, wireless adversarial interference presents high-risk challenges for the deployment of deep learning models in autonomous signal classification receivers \cite{hurdle1}.% \cite{hurdle1,hurdle2}.

% Despite their robust AMC performance, however, deep learning models are highly susceptible to gradient-based adversarial evasion attacks \cite{adv_filters,amc_adv_atk1,amc_adv_atk2,amc_adv_atk4}. Such attacks introduce additive wireless interference into transmitted RF signals to induce high-confidence misclassifications on well-trained spectrum sensing DL models. The susceptibility of DL-based AMC receivers to evasion attacks inhibits privacy and security in cognitive radios and furthermore, has been found to be more efficient than traditional jamming attacks applied in communication networks \cite{jamming}. As a result, adversarial interference in signal classification receivers presents high-risk challenges for pre-existing deep learning utilization in autonomous wireless channels \cite{hurdle1,hurdle2}.

Adversarial attacks can vary in potency depending on the amount of system knowledge available to the attacker. The most effective attack is crafted in a \emph{white box} threat model, where the adversary has knowledge of both the signal features used for classification as well as the underlying classification model (including its hyper-parameter settings and values). Not only is the availability of such specific information to an adversary rare in the context of wireless communications \cite{bb_justification}, but it is also not necessary to craft an effective adversarial interference signal. This is due to the transferability property of adversarial attacks between classification models \cite{transfer1,transfer2,transfer3}, in that an attack crafted to fool a specific DL classifier can significantly degrade performance on a disparate model trained to perform the same task. Such \emph{black box} attacks are more realistic to consider than \emph{white box} attacks in real-world communication channels, where an adversary operates with limited system information. As a result of the transferability property of adversarial attacks, an adversary can induce large-scale AMC performance degradation, thus reducing spectrum efficiency and compromising secure communication channels.

In this work, we develop a novel wireless receiver architecture capable of mitigating the effects of transferable AMC adversarial interference injected into transmitted signals in black box environments. We train a series of deep learning-based AMC models, which utilize two different representations of wireless signals as input features: (i) in-phase and quadrature (IQ) time domain signals, and (ii) frequency domain signals. Our methodology incorporates two key insights from this analysis. First, we find that \textit{attacks on DL-based AMC models are not directly transferable between signal domains} (i.e., from the time domain to the frequency domain, and vice versa). Second, we find that \textit{the transferability property varies substantially between DL architectures}. Based on these insights, we propose an ensemble AMC classification methodology that utilizes different signal representations and DL models, which we find significantly mitigates the effects of additive black box adversarial interference instantiated on both time domain and frequency domain feature sets. 

\textbf{Outline and Summary of Contributions:} Compared to related work in AMC (Sec. II), we make the following contributions:
% our defense is not attack dependent 
\begin{enumerate}
    \item \textbf{Cross-domain signal receiver architecture for AMC} (Sec. III-A, III-B, III-C, and IV-B): We develop a robust AMC module consisting of both IQ-based and frequency-based deep learning models. We show that these models obtain high classification accuracy on two real-world datasets. % We quantify the robustness of four distinct deep learning architectures for AMC using frequency-based features of RF signals. 
    
    \item \textbf{Resilience to transferable adversarial attacks between classification architectures} (Sec. III-D and Sec IV-C): We demonstrate our receiver's ability to withstand transferable adversarial interference between deep learning classification architectures trained on the same input signal representation. % We demonstrate the resilience of frequency domain feature-based AMC classifiers to time domain instantiated adversarial attacks.  % given the set of constraints placed on the adversary as described in Sec y.  
    
    %\item \textbf{Resilience to frequency domain adversaries} (Sec. xy): Similarly, we demonstrate the resilience of time domain feature-based AMC classifiers to frequency domain instantiated adversarial attacks. 
    
    \item \textbf{Resilience to transferable adversarial interference across domains} (Sec. III-E and Sec. IV-D): For a given type of DL classification architecture, we demonstrate the resilience of frequency domain trained classifiers to time domain instantiated attacks, and vice versa. This analysis leads to the identification of the most robust deep learning architectures suitable for withstanding transferable adversarial attacks.
    
    % \item \textbf{Most resilient AMC architecture to adversarial interference} (Sec. IV-D): We identify the most robust deep learning architectures suitable for withstanding transferable adversarial attacks. %Our results show that, out of several deep learning architechures, convolutional neural networks (CNNs) have the fastest training times and mitigate the classifier degradation to the greatest extent. 
    
    \item \textbf{Black box adversarial interference mitigation via deep ensemble} (Sec. III-F and Sec. IV-E): Using the foregoing properties, we develop a deep ensemble consisting of both time domain and frequency domain-based AMC classifiers trained on a variety of architectures. Our experiments show that this ensemble effectively mitigates evasion attacks regardless of the signal features or classification architecture targeted by the adversary.   
    
    % \item \textbf{Detection of adversarial interference using statistical properties of deep ensemble} (Sec. III-G and Sec. IV-F): We use our deep ensemble defense to detect high-powered adversarial attacks in cases when the perturbation masks the signal, thus degrading the ensemble's defense capabilities. 

    %Lastly, we identify the most robust deep learning architecture, which increases the classification performance on perturbed RF signals to the greatest extent.  
\end{enumerate}

\section{Related Work}
Deep learning has been widely proposed for AMC as it requires little to no feature selection to attain high classification performance on IQ samples. In particular, several studies have demonstrated the success of convolutional neural networks (CNNs) for AMC  \cite{amc_dl2,amc_dl3,amc_dl4,amc_dl5,vt_cnn2_3,ex_clf} using network graphs such as AlexNet \cite{alexnet} and ResNet \cite{resnet}. In addition to CNNs, recurrent neural networks (RNNs) have also been shown to provide high AMC accuracy \cite{rnn1,rnn2,tccn_lstm}. In this work, we build upon the success of prior AMC deep learning by proposing a series of models consisting of convolutional, recurrent, both convolutional and recurrent, and dense fully connected layers to construct the classifiers contained in our wireless signal receiver. Moreover, building on prior work \cite{fft_features,fft2}, we consider how varying the signal domain representation (time or frequency) of the input signal can impact AMC. %We demonstrate that our proposed models obtain high AMC classification performance, and further, we find that the best choice of model and input domain varies based on the dataset. 

Although the susceptibility of deep learning AMC classifiers to evasion attacks has been demonstrated in prior work \cite{adv_filters,amc_adv_atk1,amc_adv_atk2,amc_adv_atk4,def2}, relatively few defenses have been proposed to detect \cite{adv_amc_detection} or mitigate the effects of adversarial interference \cite{defense_review,def2}.  The defense algorithms which have been proposed for AMC DL classifiers -- adversarial training \cite{pgd_adv_trn,ciss_adv_trn}, Gaussian smoothing \cite{gaus_smth,kim2020channel}, and autoencoder pre-training \cite{autoencoder_defense} -- have each demonstrated degraded performance in black box environments. This is largely due to these defenses being specifically designed to defend white box attacks and being directly adopted in black box environments without special consideration being given to the differing threat model. Our proposed wireless receiver architecture, on the other hand, is designed with the intention of mitigating black box adversarial interference attacks under different knowledge levels of the adversary such as DL architecture uncertainty and classification domain uncertainty. In this regard, and to the best of our knowledge, no work has explored the extent to which adversarial attacks in AMC are transferable between signal domains (although various domains for classification have been investigated \cite{fft_features,fft2}).

Contrary to the limited defenses that exist for defending black box AMC adversarial interference, several defenses have been proposed for defending deep learning image classifiers from black box adversarial attacks, with no method generally accepted as a robust solution \cite{review1}. Nonetheless, even considering the adoption of image classification defenses for AMC is difficult due to the differing constraints placed on the adversary in each setting (e.g., channel effects and transmit power budget in AMC versus visual perceptibly and targeted pixel attacks in image classification). Therefore, in this work, we develop an ensemble defense specifically tailored to defend AMC models from adversarial attacks when the adversary is constrained by communications-based limitations. Future work may consider the adaptation of our proposed method for AMC in the image classification setting.

\section{Our AMC Methodology}

\begin{figure*}[htb] % [h] forces the figure to be output where it is defined in the code (it suppresses floating)
	\centering
	\includegraphics[width=\fullwidth\columnwidth]{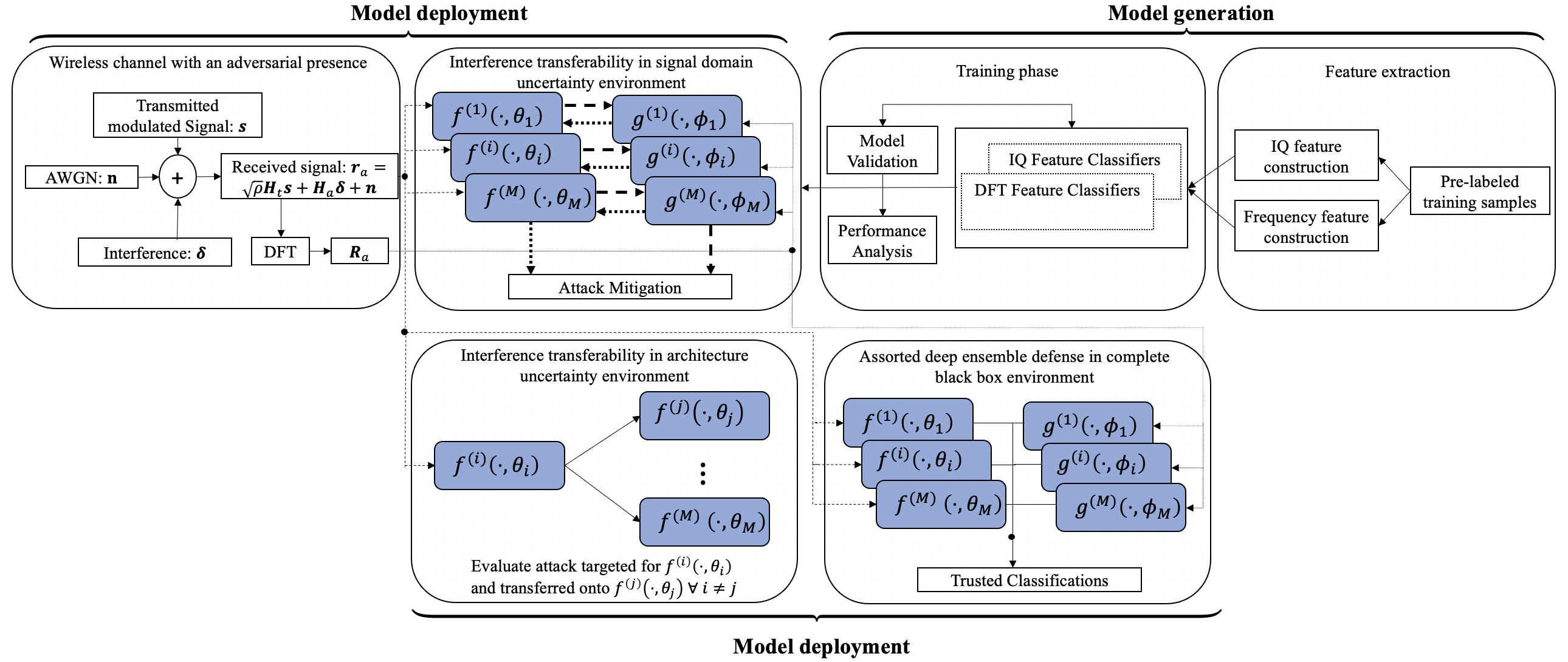}%fig1_final.png}
	% \caption{Our AMC system model with adversarial interference. The shaded blocks correspond to our constructed models deployed in the wireless receiver. The models that exhibit the highest degree of attack mitigation in both the signal domain uncertainty environment and the architecture uncertainty environment are utilized in the construction of the assorted deep ensemble defense.}
	\caption{Our AMC system model with adversarial interference. The shaded blocks correspond to our constructed models deployed in the wireless receiver, and the solid and dotted lines correspond to the adversarial frequency feature-based signal and adversarial IQ feature-based signal, respectively. The models that exhibit the highest degree of attack mitigation in both the signal domain uncertainty environment and the architecture uncertainty environment are utilized in the construction of the assorted deep ensemble (ADE) defense, whose outputs are aggregated to calculate the predicted constellation of the input signal.}
	\label{block_diagram}
\end{figure*}

In this section, we outline the wireless channel input-output model we consider for AMC as well as our proposed defense mechanisms to mitigate black box adversarial interference. We begin by describing our system model (Sec. III-A and III-B), followed by the DL classifiers that we consider for AMC (Sec. III-C). Then, we characterize the adversary's attack strategy (Sec. III-D) and define transferability relative to model architectures and signal domains (Sec. III-E). Finally, we present our deep ensemble defense for robustness against black box attacks (Sec. III-F). An overview of our methodology is given in Fig. \ref{block_diagram}. %and AMC receiver architecture (Sec. III-B) followed by the characterization of the adversary's attack strategy (Sec. III-C). Next, we develop our defense methodology in cases when (i) the adversary is blind to the classification model at the receiver (Sec. III-D), (ii) the adversary is blind to the signal's domain features used for classification, (Sec. III-E), and (iii) the adversary is blind to both the classification model and signal domain features at the receiver (Sec. III-F). Our overall AMC system model is shown in Fig. \ref{block_diagram}. 

\subsection{Signal and Channel Modeling}

We consider a wireless channel consisting of a transmitter, which is aiming to send a modulated signal, and a receiver, whose objective is to perform AMC on the received waveform and realize its modulation constellation. In addition, we consider an adversary aiming to inject interference into the transmitted signal to induce misclassification at the receiver. We will denote the channel from the transmitter to the receiver as $\mathbf{h}_{\text{t}} \in \mathbb{C}^{\ell}$ and the channel from the adversary to the receiver as $\mathbf{h}_{\text{a}} \in \mathbb{C}^{\ell}$ , where $\mathbf{h}_{\text{t}} = [h_{\text{t}}[0],\ldots,h_{\text{t}}[\ell - 1]]^{T}$, $\mathbf{h}_{\text{a}} = [h_{\text{a}}[0],\ldots,h_{a}[\ell - 1]]^{T}$, and $\ell$ denotes the length of the received signal's observation window. Both $\mathbf{h}_{\text{t}}$ and $\mathbf{h}_{\text{a}}$ also include radio imperfections such as sample rate offset (SRO), center frequency offset (CFO), and selective fading, none of which are known to the receiver. We further assume that the receiver has no knowledge of the channel model or its distribution. Therefore, the channel model is not directly utilized in the development of our methodology. However, this general setting motivates an AMC solution using a data-driven approach, as we consider in this work, in which the true modulation constellation of the received signal is estimated from a model trained on a collection of pre-existing labeled signals, which capture the effects of the considered wireless channel.

At the transmitter, we denote the transmitted signal as $\mathbf{s} = [s[0],\ldots,s[\ell - 1]]$, which is modulated using one of $C = |\mathcal{S}|$ modulation constellations chosen from a set, $\mathcal{S}$, of possible modulation schemes, with each scheme having equal probability of selection. At the receiver, the collected waveform is modeled by
\begin{equation} \label{r_t}
    \mathbf{r}_{\text{t}} =  \sqrt{\rho}\hspace{0.5mm}\mathbf{H}_{\text{t}}\mathbf{s} + \mathbf{n}
\end{equation} 
\noindent when the adversary does not instantiate an attack and as
\begin{equation} \label{r_a}
    \mathbf{r}_{\text{a}} = \mathbf{r}_{\text{t}} + \mathbf{H}_{\text{a}}\mathbf{\pmb{\delta}} =  \sqrt{\rho}\hspace{0.5mm}\mathbf{H}_{\text{t}}\mathbf{s} + \mathbf{H}_{\text{a}}\mathbf{\pmb{\delta}} + \mathbf{n}
\end{equation} 
\noindent when the adversary launches an adversarial interference signal, which is denoted by $\pmb{\delta} \in \mathbb{C}^{\ell}$, whose potency (i.e., effectiveness in inducing misclassification) is dependent on the adversary's power budget, denoted as $P_{T}$. In both (\ref{r_t}) and (\ref{r_a}), $\mathbf{r}_{\text{t}} = [r_{\text{t}}[0],\ldots,r_{\text{t}}[\ell - 1]]^{T}$ and $\mathbf{r}_{\text{a}} = [r_{\text{a}}[0],\ldots,r_{\text{a}}[\ell - 1]]^{T}$ denote the received signal in the absence and presence of adversarial interference, respectively, $\mathbf{H}_{t} = \text{diag}\{h_{\text{t}}[0],\ldots,h_{\text{t}}[\ell - 1]\} \in \mathbb{C}^{\ell \times \ell}$, $\mathbf{H}_{\text{a}} = \text{diag}\{h_{\text{a}}[0],\ldots,h_{\text{a}}[\ell - 1]\} \in \mathbb{C}^{\ell \times \ell}$, $\mathbf{n} \in \mathbb{C}^{\ell}$ represents complex additive white Gaussian noise (AWGN) distributed as $\mathcal{CN}(0,1)$, and $\rho$ denotes the signal to noise ratio (SNR), which is known at the receiver. Note that although  $\mathbf{r}_{\text{t}} = \mathbf{r}_{\text{a}}$ when $\pmb{\delta} = 0$, we define both signals separately to characterize the construction of $\pmb{\delta}$ throughout this work. 

\subsection{Signal Domain Transform} %defense using domain transform 
% take away: adversarial attacks are significantly mitigated between domains. Although some works have stated adv attacks are features in neural networks [], they are not consistent between models trained to do the same task (even the same training data is used). We exploit this to create strong defenses. 

At the receiver, we model $\mathbf{r}_{t} = [r_{t}[0],\ldots,r_{t}[\ell - 1]]^{T}$ using both (i) its in-phase and quadrature (IQ) components in the time domain and (ii) its frequency components obtained from the discrete Fourier transform (DFT) of $\mathbf{r}_{\text{t}}$. Specifically, the $p^{\text{th}}$ component of the DFT of $\mathbf{r}_{\text{t}}$ is given by 
\begin{equation} \label{dft}
    R_{t}[p] = \sum_{k=0}^{\ell - 1} r_{\text{t}}[k] e^{-\frac{j2\pi}{\ell}pk}, \hspace{2mm} p = 0,\ldots, \ell - 1, 
\end{equation}
\noindent where $\mathbf{R}_{\text{t}} = [R_{\text{t}}[0],\ldots,R_{\text{t}}[\ell - 1]]^T$ contains all frequency components of $\mathbf{r}_{\text{t}}$. Here, we are interested in comparing the effects of $\pmb{\delta}$ when an attack is instantiated on an AMC model trained on one domain (i.e., $\mathbf{r}_a$ or $\mathbf{R}_a$) and transferred to an AMC model trained on the other (i.e., $\mathbf{R}_a$ or $\mathbf{r}_a$, respectively). %either $\mathbf{r}_{a}$ or $\mathbf{R}_{a}$ and transformed into $\mathbf{R}_{a}$ or $\mathbf{r}_{a}$, respectively. 

Although both signal representations are complex (i.e., $\mathbf{r}_{t}, \mathbf{R}_{t} \in \mathbb{C}^{\ell}$), we represent all signals as two-dimensional reals, using the real and imaginary components for the first and second dimension, respectively, in order to utilize all signal components during classification. Thus, we represent all time and frequency domain features as real-valued matrices (i.e., $\mathbf{r}_{t}, \mathbf{R}_{t} \in \mathbb{R}^{\ell \times 2}$). 

%Similar to $f^{(i)}(\cdot, \theta_{i})$, we denote a deep learning classifier trained using the DFT of the input signal, $\mathbf{R}_{\text{t}}$, parameterized by $\phi_{i}$, as $g^{(i)}(\cdot, \phi_{i}): \mathbb{C}^{\ell} \rightarrow \mathbb{R}^{C}$, which is trained to perform the same classification task as $f^{(i)}(\cdot, \theta_{i})$ but using the frequency features of $\mathbf{r}_{\text{t}}$ to comprise the input signal. Then, we use $g^{(i)}(\cdot, \phi_{i})$ as a defense against adversarial attacks crafted on $f^{(i)}(\cdot, \theta_{i})$ and transferred to $g^{(i)}(\cdot, \phi_{i})$ and similarly for attacks crafted on $g^{(i)}(\cdot, \phi_{i})$ and transferred to $f^{(i)}(\cdot, \theta_{i})$. %For consistency, we use the same architectures for $g^{(i)}(\cdot, \phi_{i})$ that were used for $f^{(i)}(\cdot, \theta_{i})$. 

%We motivate the use of the DFT by assuming that the same information is represented by both $\mathbf{x}$ and $\mathbf{X}$ but the salient features of each signal differ, thus weakening adversarial perturbations crafted in the alternate domain as classification. 

\subsection{Deep Learning Architectures}

At the receiver, we consider four distinct deep learning architectures for AMC. Each considered classifier is trained on a set of modulated data signals, $\mathcal{X} \subset \mathbb{R}^{\ell \times 2}$, where each input, $\mathbf{r}_{\text{t}} \in \mathcal{X}$, belongs to one of $C$ modulation constellations and is constructed using time domain IQ features. In general, we denote a time domain trained deep learning classifier, parameterized by $\theta_{i}$, as $f^{(i)}(\cdot, \theta_{i}): \mathcal{X} \rightarrow \mathbb{R}^{C}$, where $f^{(i)}(\cdot, \theta_{i})$ for $i=1,\ldots,4$ refers to one of four deep learning architectures used to construct the classifier along with its corresponding parameters $\theta_{i}$. The trained classifier assigns each input $\mathbf{r}_{\text{t}} \in \mathcal{X}$ a label denoted by $\hat{\mathcal{C}}(\mathbf{r}_{\text{t}}, \theta_{i}) = \text{argmax}_{k} \hspace{0.5mm} f_{k}^{(i)}(\mathbf{r}_{\text{t}}, \theta_{i})$, where $f_{k}^{(i)}(\mathbf{r}_{\text{t}}, \theta_{i})$ is the vector of predicted classification probabilities, assigned by the $i^{\text{th}}$ model, of $\mathbf{r}_{\text{t}}$ being modulated according to the $k^{\text{th}}$ constellation for $k = 1,\ldots,C$. Similarly, we denote the $i^{\text{th}}$ deep learning classifier trained using the DFT of the input signal, $\mathbf{R}_{\text{t}}$, parameterized by $\phi_{i}$, as $g^{(i)}(\cdot, \phi_{i}): \mathbb{R}^{\ell \times 2} \rightarrow \mathbb{R}^{C}$, which is trained to perform the same classification task as $f^{(i)}(\cdot, \theta_{i})$ but using the frequency features of $\mathbf{r}_{\text{t}}$ to comprise the input signal. Note that $\theta_{i}$ denotes the parameters of a time domain trained classifier whereas $\phi_{i}$ denotes the parameters of frequency domain trained classifier.  

We analyze the classification performance and the efficacy of our proposed defense on four common AMC deep learning architectures: the fully connected neural network (FCNN), the convolutional neural network (CNN), the recurrent neural network (RNN), and the convolutional recurrent neural network (CRNN). For each model, we apply the ReLU non-linearity activation function in its hidden layers, given by $\nu(a) = \max\{0, a\}$, and a $C$-unit softmax activation function at the output layer given by
\begin{equation}
    \nu(\mathbf{a})_{k} = \frac{e^{a_{k}}}{\sum\limits_{j=1}^{C} e^{a_{j}}},
\end{equation}
where $k = 1,\ldots,C$ for input vector $\mathbf{a}$. This normalization allows a probabilistic interpretation of the model's output predictions. 

\textbf{FCNN}: FCNNs consist of multiple layers, which are comprised of individual units. Each unit contains a set of trainable weights, whose dimensionality is equal to the number of units in the preceding layer, and the number of units in each layer is an adjustable hyper-parameter. The output of a single unit, $u$, in a particular layer is given by
\begin{equation}
    {\nu}\bigg{(}\sum_{i} w_{i}^{(u)} a_i + b^{(u)}\bigg{)},
\end{equation}
\noindent where $\nu(\cdot)$ is the activation function, $\mathbf{w} = [w_1^{(u)}, \ldots, w_n^{(u)}]$ is the weight vector for unit $u$ estimated from the training data, $\mathbf{a} = [a_1, \ldots, a_n]$ is the vector containing the outputs from the previous layer, and $b^{(u)}$ is a threshold bias for unit $u$. 
%The FCNN input layer consists of $2 \cdot \ell$ units, where the input to each layer, $a_{i}$, is the $i^{th}$ element from the received signal, which is reshaped into a vector, i.e., $\mathbf{r}_t, \mathbf{R}_t \in \mathbb{R}^{2 \cdot \ell}$. The remainder of 
Our FCNN consists of three hidden layers with 256, 128, and 128 units, respectively, and each hidden layer applies a 20\% dropout rate during training.

\textbf{CNN}: CNNs consist of one or more convolutional layers, which extract spatially correlated patterns from their inputs. Each convolutional layer is comprised of a set of $L \times W$ filters, with the denoted as $\mathbf{m} \in \mathbb{R}^{L \times W}$. The output of the $p^{\text{th}}$ convolutional unit in a particular layer (termed a feature map) is given by
\begin{equation}
    y_{p}[j, k] = \nu\bigg{(}\sum_{l = 0}^{L-1} \sum_{w = 0}^{W-1} x[j+l, k+w] m[l^{(p)}, w^{(w)}]\bigg{)}, 
\end{equation}
\noindent where the two dimensional input, $\mathbf{x}$, and the $p^{\text{th}}$ filter, with each kernel index denoted by $m[l^{(p)}, w^{(w)}]$, are cross-correlated and passed through an activation function, $\nu(\cdot)$, to produce the $p^{\text{th}}$ feature map, $\mathbf{y}$, indexed at $j$ and $k$. 
%The input to our CNN is either $\mathbf{r}_t \in \mathbb{R}^{\ell \times 2}$ or $\mathbf{R}_t \in \mathbb{R}^{\ell \times 2}$ followed 
Our CNN is comprised of two convolutional layers consisting of 256 $2 \times 5$ and 64 $1 \times 3$ feature maps (each with 20\% dropout), respectively, and a 128-unit ReLU fully connected layer. 

\textbf{RNN}: RNNs implement feedback layers, which extract temporally correlated patterns from their inputs. Long-short-term-memory (LSTM) cells \cite{lstm} extend recurrence to create memory in neural networks by introducing three gates for learning. \emph{Input gates} prevent irrelevant features from entering the recurrent layer while \emph{forget gates} eliminate irrelevant features altogether. \emph{Output gates} produce the LSTM layer output, which is inputted into the subsequent network layer. The gates are used to recursively calculate the internal state of the cell, denoted by $\mathbf{z}_{c}^{(t)}$ at time $t$ for cell $c$, at a specific recursive iteration, called a time instance, which is then used to calculate the cell output given by
\begin{equation}
    \mathbf{q}^{(t)} = \text{tanh}(\mathbf{z}_{c}^{(t)})\nu(\mathbf{p}^{(t)}), 
\end{equation}
where $\mathbf{p}^{(t)}$ is the parameter obtained from the output gate and $\nu(\cdot)$ is the logistic sigmoid function given by $\nu(p_{i}^{(t)}) = 1 / (1 + e^{-p_{i}^{(t)}})$ for the $i^{\text{th}}$ element in $\mathbf{p}^{(t)}$. 
%Our RNN is comprised of either $\mathbf{r}_t \in \mathbb{R}^{\ell \times 2}$ or $\mathbf{R}_t \in \mathbb{R}^{\ell \times 2}$ as input followed by a 75-unit LSTM layer and a 128-unit ReLU fully connected layer.
Our RNN is comprised of a 75-unit LSTM layer followed by a 128-unit ReLU fully connected layer.

\textbf{CRNN}: Lastly, we consider a CRNN, which captures both spatial (convolutional) and temporal (recurrent) correlations in the input sequence. % Our CRNN is consists of the input ($\mathbf{r}_t \in \mathbb{R}^{\ell \times 2}$ or $\mathbf{R}_t \in \mathbb{R}^{\ell \times 2}$) followed by two convolutional layers (containing 256 $2 \times 5$ and 128 $1 \times 4$ feature maps, respectively) followed by a 128-unit LSTM layer and a 64 unit fully connected layer.
Our CRNN is consists of two convolutional layers (containing 256 $2 \times 5$ and 128 $1 \times 4$ feature maps, respectively) followed by a 128-unit LSTM layer and a 64-unit ReLU fully connected layer.

\textbf{Training Details}: Each AMC classifier contained in the wireless receiver is trained using the Adam optimizer \cite{adam} with a batch size of 64 and a dynamic learning rate scheduler. Furthermore, an early stopping criterion with a patience of 50 epochs was used as the stopping condition for training on each model to achieve convergence (i.e., the model training is terminated when the loss on a validation set has not decreased in the last 50 successive epochs). Each model was trained using the categorical cross entropy cost function, which, for the time domain feature-based models, is given by 
% Furthermore, an early stopping criterion with a patience of 50 epochs was used as the stopping condition for training on each model to achieve convergence.
% Furthermore, each classifier employed a validation set, which was used to monitor the training performance and used as the stopping condition for training on each model to achieve convergence.
\begin{equation} \label{single_cost}
    \mathcal{L}_{n}(\mathbf{r}_{t, n}, \mathbf{y}_{n}, \theta) = \sum_{j=1}^{C} y_{j} \text{log}(\hat{y}_{j})
\end{equation}
for the $n^{\text{th}}$ raining sample $\mathbf{r}_{t,n}$, and
\begin{equation} \label{total_cost}
    \mathcal{L} = -\frac{1}{N}\sum_{n=1}^{N} \mathcal{L}_{n},
\end{equation}
\noindent over the entire training set containing $N$ samples. Note that $N$ denotes the number of samples in the training set whereas $\ell$ denotes the length of the received signal's observation window. Here, $\mathbf{y}_{n}$ is the one-hot encoded label of the $n^{\text{th}}$ signal (i.e., $y_j = 1$ if the ground truth label of the sample is modulation class $j$ and $y_j = 0$ otherwise), and $\hat{y}_{j}$ is the confidence assigned by the classifier, parameterized by $\theta$, that the given input is modulated according to constellation $j$. The categorical cross entropy cost for the frequency domain feature-based models is calculated similarly to (\ref{single_cost}) and (\ref{total_cost}), but with $\mathbf{R}_{t, n}$ and $\phi$ replacing the $\mathbf{r}_{t, n}$ and $\theta$, respectively, in (\ref{single_cost}).

\subsection{Adversarial Interference}
% state for any feature representation, the attack is constructed using... 
% our defense is not attack dependent but we choose these attacks to demonstrate the effectiveness of our method 

Adversarial interference (i.e., $\pmb{\delta} \in \mathbb{R}^{\ell \times 2}$) is specifically crafted to induce erroneous modulation constellation predictions at the receiver. 
%We will denote the adversarial perturbation injected into a transmitted signal as $\pmb{\delta}: \pmb{\delta} \in \mathbb{C}^{\ell}$, where $\pmb{\delta} = ||\mathbf{x} - \tilde{\mathbf{x}}||_{p}$. We follow \cite{amc_adv_atk1} and consider an interference signal added directly to the received signal as the case when an adversary simultaneously transmits $\pmb{\delta}$ alongside $\mathbf{s}$. For a given design of $\pmb{\delta}$, the resulting signal that arrives at the receiver will be
% \begin{equation}
%     \tilde{\mathbf{x}} = \mathbf{x} + \pmb{\delta},
% \end{equation}
%where $\tilde{\mathbf{x}} = \mathbf{x}$ in the absence of an attack (i.e., when $\pmb{\delta} = 0$). %We consider a limited knowledge level threat model where the adversary knows the architecture and parameters of $f(\cdot, \theta)$ but is blind to $g(\cdot, \phi)$. This constraint mimics a real-world wireless channel where an adversary may not have complete knowledge of the underlying system under attack and thus restricts the adversary to injecting an attack in the time domain, where traditional AMC features are constructed from. %However, as noted in prior work \cite{transfer2}, adversarial attacks are transferable between classifiers so instantiating an attack on $f(\cdot, \theta)$ is expected to simultaneously degrade the performance of $g(\cdot, \phi)$, since both models are trained to solve the same classification task. 
Several methods exist to craft adversarial interference signals and, therefore, several designs of $\pmb{\delta}$ may effectively induce misclassification (i.e., disparate and unique constructions for $\pmb{\delta}$ may fulfill an adversary's objective). For a general classifier trained on IQ features, adversarial interference is crafted by solving 
\begin{subequations} \label{adv_opt:all-lines}
\begin{align}
    \underset{\pmb{\delta}}{\text{min}} \quad &  ||\pmb{\delta}||_{2} \label{adv_opt:line_1} \\
    \text{s. t.} \quad &  \hspace{0.5mm} \hat{\mathcal{C}}(\mathbf{r}_{\text{t}}, \theta) \neq \hat{\mathcal{C}}(\mathbf{r}_{\text{t}} + \mathbf{H}_{\text{a}}\pmb{\delta}, \theta), \label{adv_opt:line_2} \\ 
    \quad & ||\pmb{\delta}||_{2}^{2} \leq P_{T}, \label{adv_opt:line_3} \\
    % \quad & \mathbf{r}_{\text{t}} + \mathbf{H}_{\text{a}}\pmb{\delta} \in \mathcal{X} \label{adv_opt:line4}, 
     \quad & \mathbf{r}_{\text{t}} + \mathbf{H}_{\text{a}}\pmb{\delta} \in \mathbb{R}^{\ell \times 2} \label{adv_opt:line4}, 
\end{align}
\end{subequations} 
where $||\cdot||_{2}$ refers to the $l_{2}$ norm. The constraint given in (\ref{adv_opt:line_2}) attempts to induce misclassification with respect to the parameters of one particular targeted model (trained on time domain features) while simultaneously using the least amount of power possible to evade detection caused by higher powered adversarial interference \cite{adv_det,adv_amc_detection}, thus restricting the power budget to $P_{T}$ in (\ref{adv_opt:line_3}). Finally, (\ref{adv_opt:line4}) ensures that the perturbed sample, $\mathbf{r}_{\text{a}}$, remains in the same space as $\mathbf{r}_{\text{t}}$. An adversary may also choose to instantiate an attack on a classifier trained on frequency features; in this case, the crafted perturbation is given by replacing $\mathbf{r}_{t}$ with $\mathbf{R}_{t}$ in (\ref{adv_opt:all-lines}) while utilizing the classifier parametertized by $\phi$ instead of $\theta$. 

Note that $\pmb{\delta}$ can be constrained using other $l_{p}$ norms such as $p=0$, $p=1$, or $p = \infty$, but $p=2$ is a natural choice to consider in the domain of wireless communications as it directly corresponds to the perturbation power. Furthermore, the best solution to (\ref{adv_opt:all-lines}) is not necessarily realized when $||\pmb{\delta}||_{2}$ is minimized or when $||\pmb{\delta}||_{2}^{2} = P_{T}$, as the primary objective of the adversary is to induce misclassification on $\mathbf{r}_{a}$. In addition, a solution to (\ref{adv_opt:all-lines}) is not always guaranteed to exist, and in such cases, the additive perturbation may not necessarily result in $\mathbf{r}_{a}$ being misclassified at the receiver.

In a real-world wireless communication channel, the adversary's knowledge for constructing an attack is limited, which prevents it from solving (\ref{adv_opt:all-lines}) directly. Our focus is on black box threat models in which the adversary has some or no knowledge about the classification method at the receiver. In this capacity, we consider three distinct knowledge levels for the adversary: (i) \emph{architecture uncertainty}, where the adversary is aware of the features being used for classification (i.e., IQ vs DFT features) but unaware of the DL architecture used for classification; (ii) \emph{signal domain uncertainty}, where the adversary is aware of the DL classification architecture but unaware of the features used to comprise the input signal for classification; and (iii) \emph{overall uncertainty}, where the adversary is unaware of both the classification architecture and signal features used for classification at the receiver.

Furthermore, due to the black box threat model, the optimization in (\ref{adv_opt:all-lines}) is formulated as an \emph{untargeted} adversarial attack, where the adversary aims to induce misclassification without regard to a particular incorrect constellation assigned at the receiver. Since prior work has shown that targeted adversarial examples rarely transfer with their targeted labels \cite{transfer1}, our considered black box threat model would treat both targeted and untargeted attacks crafted on a surrogate model as untargeted attacks on the underlying target classifier. Therefore, we evaluate the resilience of our wireless receiver on untargeted attacks only. 

In this work, we consider three methods for crafting adversarial perturbations: the gradient-based fast gradient sign method (FGSM) \cite{fgsm}, the gradient-based basic iterative method (BIM) \cite{bim}, and the optimization-based Carlini and Wagner (CW) $l_{2}$ attack \cite{cw}. Under the FGSM attack, the adversary exhausts its total power budget on a single step attack, whereas under the BIM and CW attacks, the adversary iteratively uses a fraction of its attack budget, resulting in a more potent attack, in comparison to the FGSM perturbation, at the cost of higher computational overhead. In the two considered gradient-based attacks, the cost of $\mathbf{r}_{a, n}$ is first linearized as $\mathcal{L}_{n}(\mathbf{r}_{t, n} + \pmb{\delta}_{n}, \mathbf{y}_{n}, \theta) \approx \mathcal{L}_{n}(\mathbf{r}_{t, n}, \mathbf{y}_{n}, \theta) + (\mathbf{H}_{\text{a}}\pmb{\delta})^{T}\mathcal{L}_{n}(\mathbf{r}_{t, n}, \mathbf{y}_{n}, \theta)$, which is maximized by setting $\mathbf{H}_{\text{a}}{\pmb{\delta}} = \epsilon \mathcal{L}_{n}(\mathbf{r}_{t, n}, \mathbf{y}_{n}, \theta)$ where $\epsilon$ is a scaling factor to satisfy the adversary's power constraint. The optimization-based CW attack, on the other hand, optimizes (\ref{adv_opt:line_1}) subject to the constraint $\hat{\mathcal{C}}(\mathbf{r}_{\text{t}} + \mathbf{H}_{\text{a}}\pmb{\delta}, \theta) = t$, where $t$ is the second highest constellation prediction of $\mathbf{r}_{\text{t}} + \mathbf{H}_{\text{a}}\pmb{\delta}$ made by the classifier paramertized by $\theta$. This is solved by replacing the highly non-linear optimization constraint with an objective function, denoted by $\iota(\cdot)$, and reformulating the initial optimization as 
\begin{equation} \label{cw_opt}
    \underset{\pmb{\delta}}{\text{min}} \hspace{2mm} ||\pmb{\delta}||_{2} + c \cdot \iota(\mathbf{r}_{\text{t}} + \mathbf{H}_{\text{a}}\pmb{\delta}), 
\end{equation} 
which is solved using stochastic gradient descent (SGD) and the smallest possible value for $c > 0$ that results in $\iota(\mathbf{r}_{\text{t}} + \mathbf{H}_{\text{a}}\pmb{\delta}) \leq 0$.

\textbf{FGSM}: In this case, for a time domain attack the adversary adds an $l_{2}$-bounded perturbation, given by 
\begin{equation}
\pmb{\delta} = \sqrt{P_{T}} \frac { \nabla_{\mathbf{r}_{\text{t}}} \mathcal{L}_{n}(\mathbf{r}_{\text{t}, n}, \mathbf{y}_{n}, \theta)} {|| \nabla_{\mathbf{r}_{\text{t}}} \mathcal{L}_{n}(\mathbf{r}_{\text{t}, n}, \mathbf{y}_{n}, \theta)||_{2}},
\end{equation}
to the transmitted signal, $\mathbf{r}_{\text{t}, n}$, in a single step exhausting the power budget, $P_{T}$. Formally, the $n^{\text{th}}$ perturbed received signal is given by 
\begin{equation} \label{l2_fgsm}
    \mathbf{r}_{\text{a, n}} = \mathbf{r}_{\text{t}, n} + \mathbf{H}_{\text{a}} \sqrt{P_{T}} \frac {\nabla_{\mathbf{r}_{\text{t}}} \mathcal{L}_{n}(\mathbf{r}_{\text{t}, n}, \mathbf{y}_{n}, \theta)} {|| \nabla_{\mathbf{r}_{\text{t}}} \mathcal{L}_{n}(\mathbf{r}_{\text{t}, n}, \mathbf{y}_{n}, \theta)||_{2}},
\end{equation}
\noindent where $\mathcal{L}$ refers to the cost function of $f(\cdot, \theta)$ in (\ref{single_cost}). Similarly, for a frequency domain attack, the additive perturbation is given by $\pmb{\delta} = \sqrt{P_{T}} \frac { \nabla_{\mathbf{R}_{\text{t}}} \mathcal{L}_{n}(\mathbf{R}_{\text{t}, n}, \mathbf{y}_{n}, \phi)} {||\nabla_{\mathbf{R}_{\text{t}}} \mathcal{L}_{n}(\mathbf{R}_{\text{t}, n}, \mathbf{y}_{n}, \phi)||_{2}}$ resulting in the $n^{\text{th}}$ perturbed received signal being
\begin{equation} \label{l2_fgsm_frq}
    \mathbf{R}_{\text{a},n} = \mathbf{R}_{\text{t}, n} + \mathbf{H}_{\text{a}} \sqrt{P_{T}} \frac { \nabla_{\mathbf{R}_{\text{t}}} \mathcal{L}_{n}(\mathbf{R}_{\text{t}, n}, \mathbf{y}_{n}, \phi)} {||\nabla_{\mathbf{R}_{\text{t}}} \mathcal{L}_{n}(\mathbf{R}_{\text{t}, n}, \mathbf{y}_{n}, \phi)||_{2}}.
\end{equation} 
Adding a perturbation in the direction of the cost function's gradient behaves as performing a step of gradient ascent, thus aiming to increase the classification error on the perturbed sample.

\textbf{BIM}: The BIM is an iterative extension of the FGSM. Specifically, in each iteration, a fraction of the total power budget,  $\alpha < P_{T}$, is added to the perturbation, and the optimal direction of attack (the direction of the gradient) is recalculated. Formally, the total perturbation, $\mathbf{\Delta} \in \mathbb{R}^{\ell \times 2}$, is initialized to zero (i.e., $\mathbf{\Delta}_{n}^{(0)} = \mathbf{0}$), and the perturbation on iteration $k + 1$ on the $n^{\text{th}}$ sample in the time domain is calculated according to
\begin{equation}
    \mathbf{\Delta}_{n}^{(k + 1)} = \mathbf{\Delta}_{n}^{(k)} + \sqrt{\alpha} \frac {\nabla_{\mathbf{r}_{\text{t}}} \mathcal{L}_{n}(\mathbf{r}^{(k)}_{\text{t}, n}, \mathbf{y}_{n}, \theta)} {||\nabla_{\mathbf{r}_{\text{t}}} \mathcal{L}_{n}(\mathbf{r}^{(k)}_{\text{t}, n}, \mathbf{y}_{n}, \theta)||_{2}}, 
\end{equation}
which results in the perturbation given by $\pmb{\delta} = \sqrt{P_{T}}\frac {\mathbf{\Delta}_{n}} {||\mathbf{\Delta}_{n}||_{2}}$. Formally, the $n^{\text{th}}$ signal perturbed using the BIM is given by 
\begin{equation} \label{l2_bim_attack_iq}
    \mathbf{r}_{\text{a}, n} = \mathbf{r}_{\text{t}, n} + \mathbf{H}_{\text{a}}\sqrt{P_{T}} \frac {\mathbf{\Delta}_{n}} {||\mathbf{\Delta}_{n}||_{2}}. 
\end{equation}
Similarly, for a frequency domain attack, the total perturbation is calculated according to 
\begin{equation}
    \mathbf{\Delta}_{n}^{(k + 1)} = \mathbf{\Delta}_{n}^{(k)} + \sqrt{\alpha} \frac {\nabla_{\mathbf{R}_{\text{t}}} \mathcal{L}_{n}(\mathbf{R}^{(k)}_{\text{t}, n}, \mathbf{y}_{n}, \phi)} {||\nabla_{\mathbf{R}_{\text{t}}} \mathcal{L}_{n}(\mathbf{R}^{(k)}_{\text{t}, n}, \mathbf{y}_{n}, \phi)||_{2}}, 
\end{equation}
yielding the perturbed frequency domain signal
\begin{equation} \label{l2_bim_attack_frq}
    \mathbf{R}_{\text{a}, n} = \mathbf{R}_{\text{t}, n} + \mathbf{H}_{\text{a}}\sqrt{P_{T}} \frac {\mathbf{\Delta}_{n}} {||\mathbf{\Delta}_{n}||_{2}}, 
\end{equation}
\noindent where the final additive perturbation, $\mathbf{\Delta}_{n}$, in both the time and frequency domain is scaled by $\frac{\sqrt{P_{T}}} {||\mathbf{\Delta}_{n}||_{2}}$ to satisfy the power constraint of the adversary.

\textbf{CW}: The CW attack seeks to optimize (\ref{cw_opt}) using SGD. Formally, the adversarial perturbation, with the effect of the adversary's channel, on the $n^{\text{th}}$ received signal is given by 
\begin{equation}
    %\mathbf{H}_{\text{a}}\mathbf{r}_{\text{a}, n} =
    \underset{\mathbf{w}}{\text{min}} \bigg|\bigg|\frac{1}{2}\bigg(\text{tanh}(\mathbf{w}) + 1\bigg) - \mathbf{r}_{t,n}\bigg|\bigg|^{2}_{2} + c \cdot \iota\bigg(\frac{1}{2}(\text{tanh}(\mathbf{w}) + 1)\bigg), 
\end{equation}
where $c$ is a constant and $\iota(\cdot)$ is empirically found to be 
% \begin{equation}
%     \iota(\mathbf{r}_{\text{a}}) = \text{max}(\text{argmax}\{Z(\mathbf{r}_{\text{a}}, \theta)_{i}: i \neq t\} - Z(\mathbf{r}_{\text{a}}, \theta)_{t}, -\kappa), 
% \end{equation}
\begin{equation}
    \iota(\mathbf{r}_{\text{a}}) = \text{max}\bigg{(}\underset{i \neq t}{\text{max}}\{Z(\mathbf{r}_{\text{a}}, \theta)_{i} - Z(\mathbf{r}_{\text{a}}, \theta)_{t}\}, -\kappa\bigg{)}, 
\end{equation}
where $Z(\cdot, \theta)$ outputs the classifier's logit vector for $(\cdot$) (i.e., the classifier predictions prior to softmax normalization), $i$ is the correct modulation prediction, and $t$ is the second most likely constellation prediction of $\mathbf{r}_{\text{a}}$ made by the classifier parameterized by $\theta$. In addition, the parameter $\kappa$ controls the confidence with which misclassification occurs (nominally to class $t$) on $\mathbf{r}_{\text{a}}$. $\kappa$ can be interpreted as a confidence parameter, with higher $\kappa$ making a sample more likely to be misclassified when transferred to a different model. More details on this derivation are given in \cite{cw}.
% put the opt and define its functions here. explain kappa, correct class, target class, and logits too. 

For each attack, we assume a naive adversarial attack instantiation, where we set $\mathbf{H}_{\text{a}} = \mathbf{I}$, following \cite{amc_adv_atk1,kim2020channel,autoencoder_defense}. As a result, each element of the crafted perturbation retains its sign and magnitude after going through the channel. This general setup focuses on controlling the model's behavior at the transmitter and receiver, while considering the most stringent threat model for the adversary. Furthermore, since $\mathbf{H}_{\text{a}} = \mathbf{I}$, $P_{T}$ directly corresponds to both the adversary's power constraint as well as the received power at the sink.

\subsection{Resilience to Transferable Interference} %Defense Using Mismatched Architecture
% take away: although adv attacks are transferable, they degrade the performance of different architectures to a lesser degree 

We demonstrate the resilience of our wireless AMC receiver to transferable adversarial interference in architecture uncertainty and signal domain uncertainty environments. In the architecture uncertainty threat model, the adversary has access to one of the classifiers contained within the receiver trained on IQ time domain samples. In this case, $\pmb{\delta}$, in both the FGSM and BIM attacks, is constructed using the gradient of the accessible classifier and transmitted alongside $\mathbf{r}_{t}$. Specifically, we evaluate the improvement provided by $\hat{\mathcal{C}}(\mathbf{r}_{\text{a}}, \theta_{j}) = \text{argmax}_{k} \hspace{0.5mm} f_{k}^{(j)}(\mathbf{r}_{\text{a}}, \theta_{j})$ when an attack is crafted using $P_{T}$ and $\nabla_{\mathbf{r}_{\text{t}}}\mathcal{L}(\mathbf{r}_{\text{t}}, \mathbf{y}, \theta_{i})$, which we evaluate $\forall i \neq j$. 

In the signal domain uncertainty environment, the adversary crafts $\pmb{\delta}$ using the gradient of either $f^{(i)}(\cdot, \theta_{i})$ or $g^{(i)}(\cdot, \phi_{i})$ and attempts to transfer the attack onto $g^{(i)}(\cdot, \phi_{i})$ or $f^{(i)}(\cdot, \theta_{i})$, respectively. Formally, the transferability of an attack from the time domain to the frequency domain is assessed through the accuracy improvement provided by  $f^{(i)}(\cdot, \theta_i)$ when the adversary operates based on $g^{(i)}(\cdot, \phi_i)$ (and vice versa for assessing the transferability of an attack from the frequency domain to the time domain). In this scenario, we demonstrate the resilience of our wireless AMC receiver to transferable attacks targeted at degrading domain specific classifiers.

\subsection{Assorted Deep Ensemble Defense} %Defense Using Assorted Deep Ensembles
% take away: (ADE) this provides the most robust defense compared to transferring from just a single model to another single model. Most DEs use the same architecture and input space for all models. Based on the advantages that certain archs have against others, we train an ensemble with a variety of archs and input reps. State as list of models and describe how it works (maybe use an algorithm) 

\begin{algorithm}[t] % use t for double column and H for single column
   \caption{ADE Construction}
   \label{ade_const}
   \begin{algorithmic}[1] % for line numbers
        \STATE \textbf{input:} $\mathcal{X}^{\text{IQ}}$: IQ feature-based training set \\ 
        \hspace{9mm} $\mathcal{X}^{\text{DFT}}$: frequency feature-based training set \\
        \hspace{9mm} $\pmb{\zeta} = \{\zeta^{(1)},\ldots, \zeta^{(M)}\}$: set of randomly initialized \\
        \hspace{9mm} untrained deep learning architectures \\
        \hspace{9mm} $k$: number of noisy samples generated per signal \\
        \hspace{9mm} $\sigma_{\text{IQ}}$: standard deviation of Gaussian noise added to \\
        \hspace{9mm} IQ samples \\
        \hspace{9mm} $\sigma_{\text{DFT}}$: standard deviation of Gaussian noise added \\
        \hspace{9mm} to DFT samples \\
        
        % initialize F and G to empty sets 
        \STATE \textbf{initialize:} $F \gets \emptyset$ \\
        \hspace{14mm} $G \gets \emptyset$ 
        
        \FOR{$i = 1, \ldots, M$} 
        
            \FOR{$j = 1, \ldots, k$}
        
                \STATE $\mathcal{X}^{\text{IQ}}_{\text{noisy}} \gets \emptyset$
                \STATE $\mathcal{X}^{\text{DFT}}_{\text{noisy}} \gets \emptyset$
                
                \FOR{$\mathbf{r}_{\text{t}}, \mathbf{R}_{\text{t}} \in \mathcal{X}^{\text{IQ}}, \mathcal{X}^{\text{DFT}}$}
                
                    \STATE $\mathbf{n}_j^{\ell \times 2} \gets \mathcal{N}(\mu=0, \sigma=\sigma_{\text{IQ}})$
                    \STATE $\mathbf{N}_j^{\ell \times 2} \gets \mathcal{N}(\mu=0, \sigma=\sigma_{\text{DFT}})$
                    % \STATE $\mathbf{n}_j^{\ell \times 2} \gets \text{RandNormal}(\mu=0, \sigma=\sigma_{\text{IQ}})$
                    % \STATE $\mathbf{N}_j^{\ell \times 2} \gets \text{RandNormal}(\mu=0, \sigma=\sigma_{\text{DFT}})$
                    \STATE $\tilde{\mathbf{r}}_{\text{t}} \gets \mathbf{r}_{\text{t}} + \mathbf{n}_j$
                    \STATE $\tilde{\mathbf{R}}_{\text{t}} \gets \mathbf{R}_{\text{t}} + \mathbf{N}_j$
                    \STATE $\mathcal{X}^{\text{IQ}}_{\text{noisy}} \gets \mathcal{X}^{\text{IQ}}_{\text{noisy}} \cup \tilde{\mathbf{r}}_{\text{t}}$ 
                    \STATE $\mathcal{X}^{\text{DFT}}_{\text{noisy}} \gets \mathcal{X}^{\text{DFT}}_{\text{noisy}} \cup \tilde{\mathbf{R}}_{\text{t}}$ 
                    % \STATE add $\tilde{\mathbf{r}}_{\text{t}}$ to $\mathcal{X}^{\text{IQ}}_{\text{noisy}}$
                    % \STATE add $\tilde{\mathbf{R}}_{\text{t}}$ to $\mathcal{X}^{\text{DFT}}_{\text{noisy}}$
            
                \ENDFOR
                
            \ENDFOR 
            
            \STATE $f^{(i)}(\cdot, \theta_{i}) \gets$ train $\zeta^{(i)}$ on $\mathcal{X}^{\text{IQ}}_{\text{noisy}}$
            \STATE $F \gets F \cup f^{(i)}(\cdot, \theta_{i})$
            % \STATE add $f^{(i)}(\cdot, \theta_{i})$ to $F$
            \STATE $g^{(i)}(\cdot, \phi_{i}) \gets$ train $\zeta^{(i)}$ on $\mathcal{X}^{\text{DFT}}_{\text{noisy}}$
            \STATE $G \gets  G \cup g^{(i)}(\cdot, \phi_{i})$
            % \STATE add $g^{(i)}(\cdot, \phi_{i})$ to $G$
        
        \ENDFOR 

        \RETURN $F, G$
  
  \end{algorithmic}

\end{algorithm}

\begin{algorithm}[t] % use t for double column and H for single column
   \caption{ADE Deployment}
   \label{ade_def}
   \begin{algorithmic}[1] % for line numbers
        \STATE \textbf{input:} $\mathbf{r}_{\text{a}}$: perturbed wireless signal \\ 
        \hspace{9mm} $F = \{f^{(1)}(\cdot, \theta_{1}),\ldots, f^{(M)}(\cdot, \theta_{M})\}$ \\ %: set of IQ classifiers \\
        \hspace{9mm} $G = \{g^{(1)}(\cdot, \phi_{1}),\ldots, g^{(M)}(\cdot, \phi_{M})\}$ \\ %: set of DFT classifiers
        
        \STATE \textbf{initialize:} $\hat{q}_{f} \gets \mathbf{0}^{M \times C}$ \\
        \hspace{14mm} $\hat{q}_{g} \gets \mathbf{0}^{M \times C}$

        \FOR{$i = 1, \ldots, M$}
            
            \STATE $\hat{q}_{f}[i] \gets f^{(i)}(\mathbf{r}_{\text{a}}, \theta_{i})$
            \STATE $\mathbf{R}_{\text{a}} \gets \sum\limits_{k=0}^{\ell - 1} {r}_{\text{a}}[k] e^{-\frac{j2\pi}{\ell}pk} \hspace{2mm} \textbf{for} \hspace{2mm} p = 0,\ldots, \ell - 1$
            \STATE $\hat{q}_{g}[i] \gets g^{(i)}(\mathbf{R}_{\text{a}}, \phi_{i})$

            \ENDFOR 
            
        \STATE $\hat{Q}^{2M \times C} \gets [\hat{q}_{f}; \hat{q}_{g}]$
        \STATE $\hat{\mathbf{y}} \gets \mathbf{0}^{C \times 1}$
        \FOR{$j = 1,\ldots,C$}
            \STATE $\hat{y}[j] \gets \frac{1}{2M} \sum\limits_{i=1}^{2M} \hat{Q}[i, j]$
        \ENDFOR 
        \STATE $\hat{C} \gets \text{argmax}_{i} \hspace{1mm} \hat{\mathbf{y}}_{i}$
        \RETURN $ \hat{C}$
  
  \end{algorithmic}

\end{algorithm}

\begin{figure*}[htb] % [h] forces the figure to be output where it is defined in the code (it suppresses floating)
	\centering
	\includegraphics[width=\fullwidth\columnwidth]{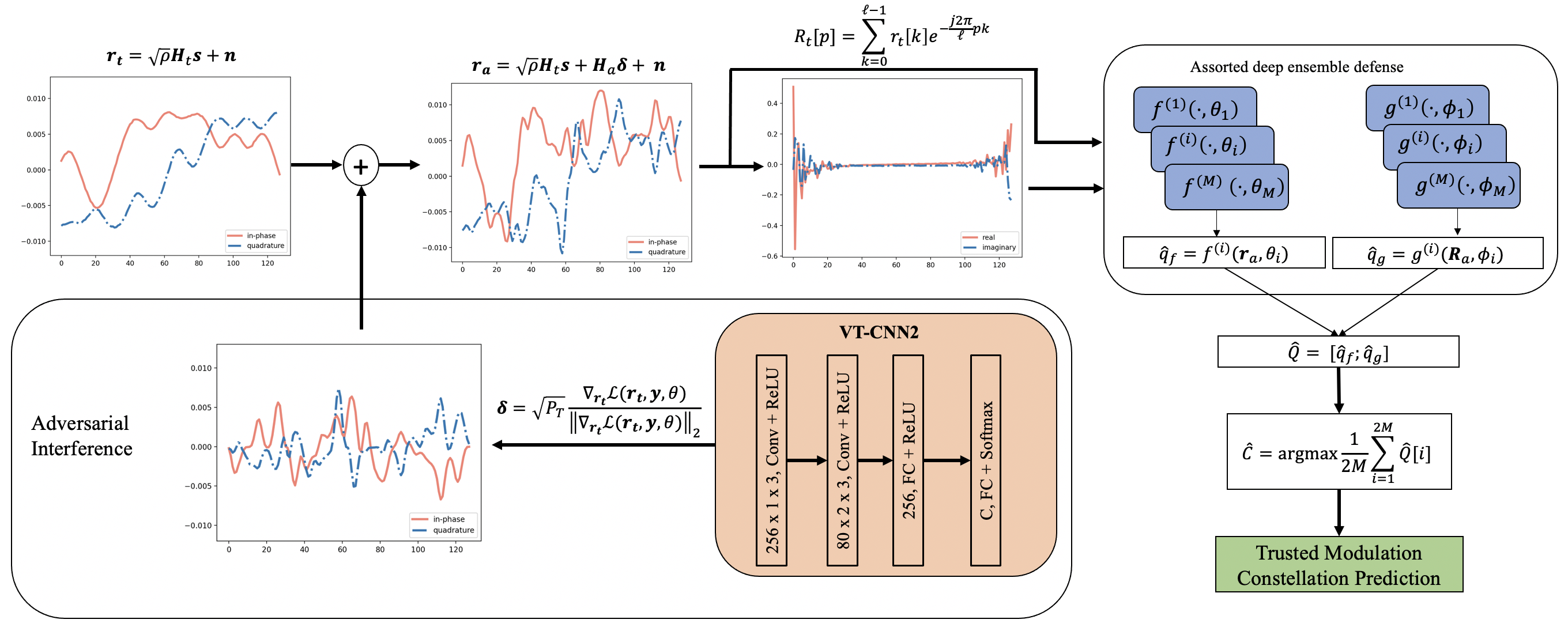}%fig1_final.png}
	\caption{Illustration of our AMC methodology in the black box adversarial attack environment on a GFSK modulation signal, where the adversary is forced to use the gradient of a surrogate model to craft the interference signal.}
	\label{bb_env}
\end{figure*}

We now develop our defense against adversarial interference in a complete black box attack environment in which the adversary is blind to both the classification architecture and the signal domain used at the receiver. Here, we introduce our assorted deep ensemble (ADE) defense, which offers diversity among both classification architectures and signal representations. Contrary to deep ensemble models that have been proposed for other applications in prior work \cite{deep_ensembles}, our proposed ADE for AMC employs a variety of models trained on both IQ and frequency-based features. Furthermore, each classifier contained in our ADE defense is trained using Gaussian smoothing, which improves classification performance on out-of-distribution waveforms. Specifically, Gaussian smoothing involves adding multiple copies of each training sample into the training set, where each copied signal is randomly perturbed. When the training dataset is augmented with a sufficient amount of random perturbations in this fashion, the classification performance of a single DL model increases on adversarial examples, since the random noise accounts for various distortions that may be induced by adversarial examples \cite{gaus_smth}. Finally, each classifier in our ADE is trained using the entire available training set (as opposed to bootsrap aggregating, i.e., \emph{bagging}, traditionally used in ensemble training), where different random initializations (as well as, in our case, Gaussian noise signal perturbations in the training set) are used to create diversity among the models. 

Algorithm \ref{ade_const} outlines the training process for our proposed ADE. Here, $\mathcal{N}$ denotes generating an $\ell \times 2$ matrix from a Gaussian distribution. Consistent with the design of black box attacks crafted in prior work \cite{transfer1,transfer2,transfer3,review1}, the adversary uses a surrogate deep learning classifier to craft their attack, which is transmitted to the underlying classification model at the receiver (we will discuss the procedure used by the adversary to construct the surrogate model in our experiments in Sec. IV-E). In this fashion, our defense is especially applicable in overall uncertainty black box environments when an adversary cannot access the gradient of the underlying classifier at the receiver. %In particular, our defense aims to mitigate the effects of additive adversarial interference crafted using the gradient of a surrogate model, accessible to the adversary, in their attempt to construct a transferable attack (we will discuss the procedure used by the adversary to construct the surrogate model in our experiments in Sec. IV-E). As a result, our defense is especially applicable in overall uncertainty black box environments when an adversary cannot access the gradient of the underlying classifier at the receiver. 

During deployment, the modulation constellation of an input, $\mathbf{r}_{\text{a}}$, is predicted by aggregating the output of each classifier in the ensemble trained on IQ features, $F=\{f^{(1)}(\cdot,\theta_{1}),\ldots,f^{(M)}(\cdot, \theta_{M})\}$, and frequency features, $G =\{g^{(1)}(\cdot, \phi_{1}),\ldots,g^{(M)}(\cdot, \phi_{M})\}$. Algorithm \ref{ade_def} outlines the deployment of our ADE defense against adversarial signals. The application of our black box defense is illustrated in Fig. \ref{bb_env}.

\section{Results and Discussion}

% In this section, we conduct an empirical evaluation of our AMC methodology. First, we overview the datasets that we use (Sec. IV-A). Next, we present the efficacy of our wireless receiver using both IQ and frequency features for classification in the absence of any adversarial interference (Sec. IV-B). We then demonstrate our wireless receiver's resilience to transferable adversarial interference between classification architectures (Sec. IV-C) and signal domains (Sec. IV-D). Finally, we evaluate our assorted deep ensemble (ADE) defense and demonstrate its robustness in black box attack environments over two comparative baselines (Sec. IV-E). 

In this section, we conduct an empirical evaluation of our AMC methodology. First, we overview the datasets that we use (Sec. IV-A). Next, we present the baseline classification performances in the absence of adversarial interference (Sec. IV-B). We then demonstrate our wireless receiver's resilience to transferable adversarial interference between classification architectures (Sec. IV-C) and signal domains (Sec. IV-D). Finally, we evaluate our assorted deep ensemble (ADE) defense and demonstrate its robustness in black box attack environments over three comparative baselines (Sec. IV-E). 

\subsection{Datasets and Evaluation Metrics}

We employ the GNU RadioML2016.10a (referred to as Dataset A) \cite{dataset} and RadioML2018.01a (referred to as Dataset B) \cite{amc_dl3} datasets for our analysis. Each signal in the RadioML2016 dataset ($\mathbf{r}_{\text{t}, n}$) is normalized to unit energy and consists of a 128-length ($\ell = 128$) observation window modulated according to a certain constellation ($\mathbf{y}_n$). In order to isolate the impact of adversarial interference, we focus on the following four modulation schemes, which, as we will show in Sec. IV-B, have equivalent classification performance on both time and frequency domain features in the absence of adversaries: CPFSK, GFSK, PAM4, and QPSK. Each constellation set contains 6000 examples for a total of 24000 signals. In the RadioML2018.01a dataset, each considered signal is oversampled with an observation window of $\ell = 1024$. For a consistent comparison between datasets, signals in the RadioML2018.01a dataset are downsampled by $1 / 8$ to obtain an observation window of $\ell = 128$. We then focus on the following modulation constellations within the dataset: OOK, 8ASK, BPSK, and FM. Each modulation scheme contains 4096 examples for a total of 16384 signals. Following in line with previous studies \cite{amc_adv_atk1,kim2020channel,autoencoder_defense}, we focus on high SNR signals, where small perturbations can significantly degrade the classification performance of otherwise robust AMC models. Therefore, for each considered dataset, we present results at a constant SNR of 18 dB unless otherwise stated. However, our conclusions hold at SNRs greater than 2 dB, where previous works (e.g., \cite{fft_features}) have achieved high classification accuracy for deep learning-based AMC. We will analyze the effects of varying the SNR more closely in Sec. IV-E.

%Moreover, we found that our method delivers nearly identical results, in comparison to the high SNR of 18 dB at which results are shown in this paper, on each considered dataset when the SNR is greater than 2 dB. At SNRs lower than 2 dB, baseline classification performance, in the absence of adversarial interference, is degraded due to poor signal quality (consistent with prior work, e.g., \cite{amc_dl1,amc_dl2,amc_dl3}). In these cases, our defense does provide gains in accuracy in the presence of adversarial interference but, due to the low signal quality, the classification performance, even of the improved gains in accuracy, are not very effective due to the poor signal environment induced from low SNR waveforms. Therefore, following in line with previous studies \cite{amc_adv_atk1,kim2020channel,autoencoder_defense}, we focus on high SNR signals, where small perturbations can significantly degrade the classification performance of otherwise robust AMC models.

In each experiment, we employ a 70/15/15 training/validation/testing dataset split, where the training and validation data are used to estimate the parameters of $f^{(i)}(\cdot, \theta_{i})$ and $g^{(i)}(\cdot, \phi_{i})$, and the testing dataset is used to evaluate each trained model's susceptibility to adversarial interference as well as the effectiveness of our proposed defense. In particular, the validation set is used to tune the model parameters using unseen data during the training process whereas the testing set is used to measure the performance of the resulting model. For each dataset, we denote the training, validation, and testing datasets, consisting of time domain IQ points or frequency domain feature components, as $\mathcal{X}_{tr}^{t}$, $\mathcal{X}_{va}^{t}$, $\mathcal{X}_{te}^{t}$, $\mathcal{X}_{tr}^{\omega}$, $\mathcal{X}_{va}^{\omega}$, and $\mathcal{X}_{te}^{\omega}$, respectively. 
% say that we only focus on high snr signals because prior work has shown Gaussian noise to interfere with adversary's objective

To measure the potency of the additive adversarial interference, we use the perturbation-to-noise ratio (PNR), which is given by
\begin{equation}
     \text{PNR} \hspace{1mm} [\text{dB}] = \frac {\mathbb{E}[\|\pmb{\delta}\|_{2}^2]}  {\mathbb{E}[\|\mathbf{r}_a\|_2^2]}  \hspace{1mm} [\text{dB}] + \text{SNR} \hspace{1mm} [\text{dB}],
    %\text{PNR} \hspace{1mm} [\text{dB}] = 10 \cdot \text{log}\bigg{(}\frac {||\pmb{\delta}||^{2}_{2}} {||\mathbf{r}_{\text{a}}||^{2}_{2}}\bigg{)} \hspace{1mm} [\text{dB}] + \text{SNR} \hspace{1mm} [\text{dB}].
\end{equation}
\noindent where $\mathbb{E}$ is the expected value. A higher PNR indicates higher levels of additive interference. We consider a perturbation to be imperceptible when $\text{PNR} \leq 1 \hspace{1mm} \text{dB}$ because the perturbation power would be at or below the noise power. At high PNR (i.e., $\text{PNR} > 1 \hspace{1mm} \text{dB}$), the underlying signal is masked to a greater extent by the perturbation, making effective classification difficult in any case due to the loss of salient features across classification architectures. 

At each PNR, we measure the accuracy of the considered testing set (i.e., $\mathcal{X}_{te}^{t}$ or $\mathcal{X}_{te}^{\omega}$), which is given by dividing the total number of correctly classified samples by the total number of samples in the set. Although random predictions would yield an accuracy of 25\% (since $C=4$) in our experiments, we will see that adversarial perturbations can result in classification accuracies significantly below random guessing, indicating their potency on DL-based wireless communication networks. %Since, in our experiments, $C = 4$, random predictions would result in a classification accuracy of 25\% on each dataset. However, as we will see, the effect of the adversarial perturbation can induce prediction accuracies significantly below random guessing, demonstrating their potency on DL-based wireless communication networks. 

% Although all of our subsequent results are shown for a fixed SNR of 18 dB, our conclusions hold at SNRs greater than 2 dB, where previous works (e.g., \cite{fft_features}) have achieved high classification accuracy for deep learning-based AMC. In fact, we find that our method delivers nearly identical results, in comparison to the high SNR of 18 dB at which the experiments are conducted in this paper, on each considered dataset when the SNR is greater than 2 dB. Thus, we omit these results for brevity. At SNRs lower than 2 dB, baseline classification performance, in the absence of adversarial interference, is degraded due to poor signal quality. In these cases, our defense does provide gains in accuracy in the presence of adversarial interference but, due to the low signal quality, the classification performance, even of the improved gains in accuracy, are not very effective due to the poor signal environment. Therefore, the effects of adversarial interference are more interesting to consider in the high SNR regime (as has been considered in previous related works \cite{amc_adv_atk1,kim2020channel,autoencoder_defense}), where small perturbations can significantly degrade the classification performance of otherwise robust AMC models. 

\subsection{AMC Wireless Receiver Performance}

%%%%%%%%%%%%%%%%%%%%%%%%

% use below for double column figures

%%%%%%%%%%%%%%%%%%%%%%%%

\begin{figure}[t] % [h] forces the figure to be output where it is defined in the code (it suppresses floating)
	\centering
	\includegraphics[width=\halfwidth\columnwidth]{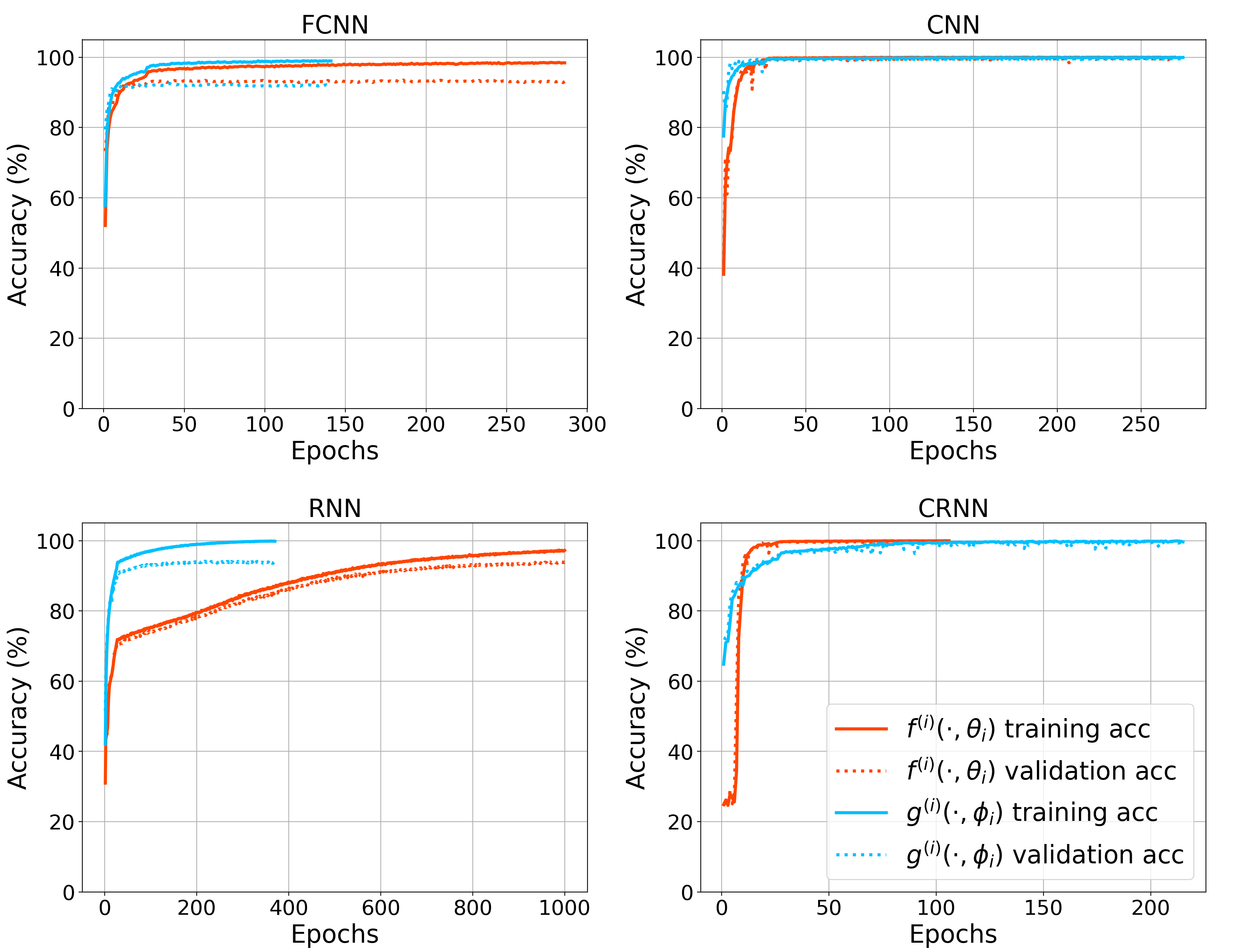}
	\caption{The model training performance on Dataset A of each considered AMC architecture on the corresponding training and validation sets. We see that the frequency-based features $g(\cdot, \phi)$ outperform or match the time domain features $f(\cdot, \theta)$ in terms of both training and validation accuracy for each deep learning architecture. The CNN results in the fastest convergence and highest accuracy for both $f(\cdot, \theta)$ and $g(\cdot, \phi)$.}
	\label{iq_frq_perf}
\end{figure}

\begin{figure}[t] % [h] forces the figure to be output where it is defined in the code (it suppresses floating)
	\centering
	\includegraphics[width=\halfwidth\columnwidth]{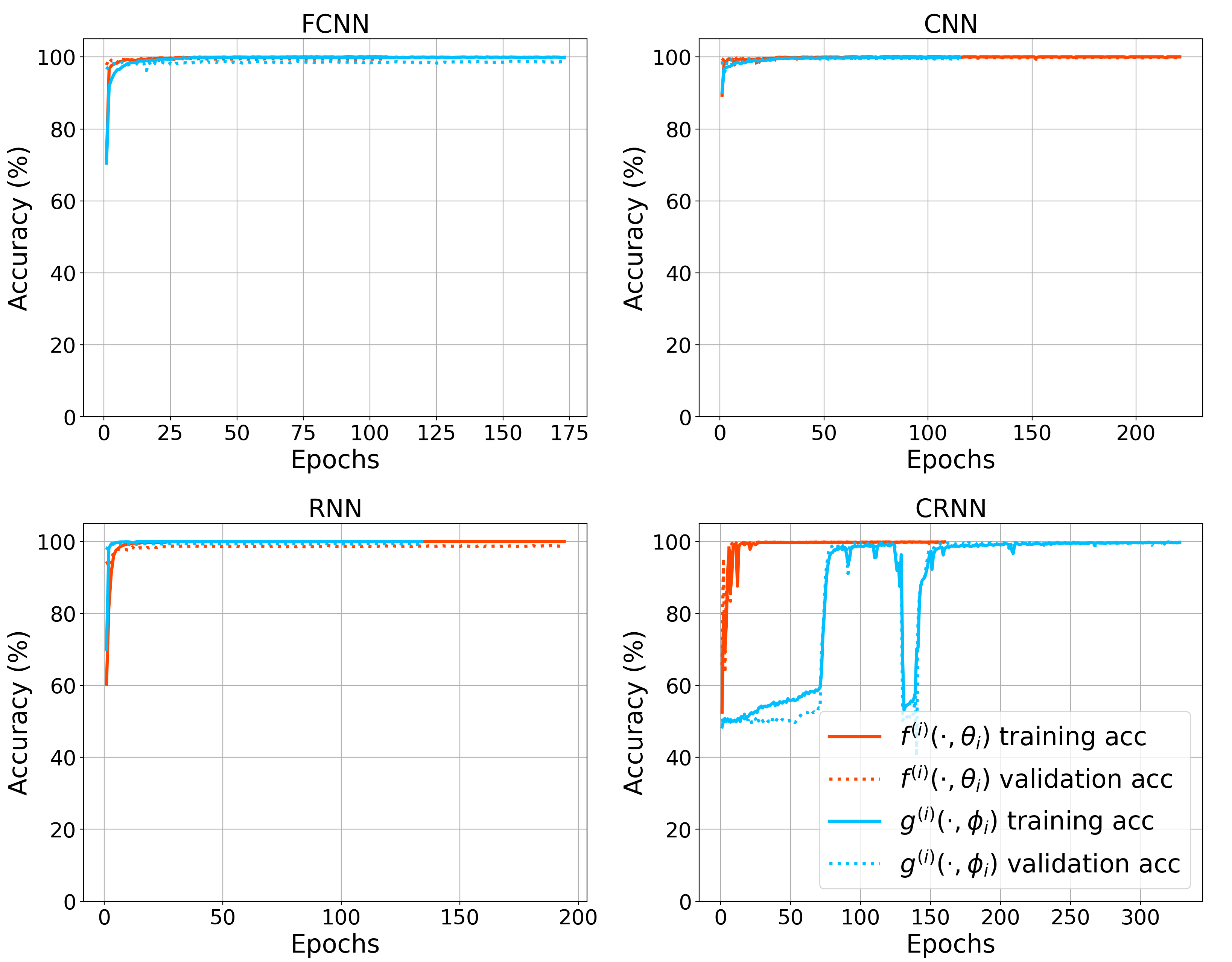}
	\caption{The model training performance on Dataset B of each considered AMC architecture on the corresponding training and validation sets. Similar to Fig. \ref{iq_frq_perf}, we see that the CNN and CRNN achieve robust classification performance on both IQ and frequency features. Unlike Fig. \ref{iq_frq_perf}, the FCNN and RNN experience similar performance to the CNN and CRNN.}
	\label{iq_frq_perf_ds2}
\end{figure}

\begin{table} [t]
\small
\caption{The testing accuracy of each considered model on $\mathcal{X}_{te}^{(\cdot)}$ as well as the Pearson Product-Moment correlation coefficient (PPMCC) between the validation sets of Dataset A and Dataset B. The CNN outperforms every other considered model (although the CRNN delivers equivalent accuracy, it is achieved with a longer training time on both datasets). \label{model_acc}}
\centering
\begin{tabular}{c c c c c} 
\centering 
\makecell{Model} & \makecell{Input \\ Features} & \makecell{Accuracy on \\ Dataset A} & \makecell{Accuracy on \\ Dataset B} & \makecell{PPMCC} \\
\hline
FCNN & IQ &  92.78\% & 99.35\% & 0.64 \\
FCNN & Frequency & 92.22\% & 99.10\% & 0.57\\
CNN & IQ & 99.19\% & 99.59\% & 1.0 \\
CNN & Frequency &  99.25\% & 99.72\% & 0.85\\
RNN & IQ &  93.61\% & 98.58\% & 0.80\\
RNN & Frequency & 92.53\% & 99.15\% & 0.65\\
CRNN & IQ & 99.61\% & 99.59\% & 0.80\\
CRNN & Frequency &  98.92\% & 99.63\% & 0.52\\

\hline

\end{tabular}
\end{table}

We begin by evaluating the performance of both $f^{(i)}(\cdot, \theta_{i})$ and $g^{(i)}(\cdot, \phi_{i})$ in the absence of adversarial interference. In Figs. \ref{iq_frq_perf} and \ref{iq_frq_perf_ds2}, we plot the evolution of the classification accuracy across training epochs achieved by each deep learning architecture on their respective training and validation sets. \emph{We see that each model trained using our proposed frequency feature-based input performs equivalently or outperforms its time domain counter-part model in terms of final accuracy (calculated with regards to the subset of considered modulation constellations in each dataset) and required training epochs, with the exception of the CRNN.} %For example, the RNN trained on frequency components in each dataset achieves higher classification performance on $\mathcal{X}_{tr}^{\omega}$ compared to $\mathcal{X}_{tr}^{t}$ in significantly fewer epochs. 
We also see in Figs. \ref{iq_frq_perf} and \ref{iq_frq_perf_ds2} that, \emph{among each considered architecture, the CNN and CRNN consistently obtain the best performance overall on their validation sets.} Contrarily, both IQ and frequency features present more challenges during training on the FCNN and RNN compared to the CNN and CRNN on Dataset A, whereas the CRNN presents more training instability on Dataset B. Specifically, on Dataset A, both the FCNN and the RNN experience slight overfitting to the training data and fail to converge on a validation accuracy greater than 94\%, while the CNN and CRNN models present generalizable and robust performance nearing or exceeding 99\%. Dataset B, however, experiences a sudden drop in training accuracy on the CRNN but does not experience overfitting on any classifier trained on either IQ or frequency features, while delivering a validation accuracy greater than 99\% on each classification architecture. % The CNN, on the other hand, entails almost no degree of overfitting while converging in substantially less epochs compared with the RNN and CRNN models. 

Each trained model's accuracy achieved on its corresponding testing set is shown in Table \ref{model_acc}. In addition, we show the Pearson Product-Moment correlation coefficient (PPMCC) \cite{ppmcc} between the validation accuracies (at the end of each training iteration) on Dataset A and Dataset B for the same set of signal features (i.e., IQ vs. FFT) on each dataset. Among all eight considered models, the CNN trained on frequency features achieves the highest testing accuracy while converging using the fewest epochs as shown in Figs. \ref{iq_frq_perf} and \ref{iq_frq_perf_ds2}. Although the CRNN achieves a slightly higher testing accuracy, the higher number of required epochs results in substantially higher computational overhead (the CNN converges three times faster than the CRNN). Moreover, the CNN has the highest PPMCC correlation between Dataset A and Dataset B on their corresponding classifiers. \emph{Therefore, our proposed CNN trained using frequency features is the most desirable model in terms of classification performance, training time, consistency between datasets, and computational efficiency.} 

% Each trained model's accuracy achieved on its corresponding testing set is shown in Table \ref{model_acc}. Among all eight considered models, the CNN trained on frequency features achieves the highest testing accuracy while converging using the fewest epochs as shown in Figs. \ref{iq_frq_perf} and \ref{iq_frq_perf_ds2}. Although the CRNN achieves a slightly higher testing accuracy, the higher number of required epochs results in substantially higher computational overhead (the CNN converges three times faster than the CRNN). \emph{Therefore, our proposed CNN trained using frequency features is the most desirable model in terms of classification performance, training time, and computational efficiency.} 

\subsection{Architecture Uncertainty Performance}

% \begin{figure}[t] % [h] forces the figure to be output where it is defined in the code (it suppresses floating)
% 	\centering
% 	\includegraphics[width=\halfwidth\columnwidth]{iq_trf_fgsm_2_plots2.png}
% 	\caption{The transferability of the FGSM perturbation between classification architectures on Dataset A. We see that, in the low PNR regime ($\text{PNR} < 0$), the effects of FCNN, CNN, and RNN instantiated attacks are nearly eliminated on the CRNN while the effects of CRNN instantiated attacks fail to strongly transfer onto RNNs.}
% 	\label{fgsm_l2}
% \end{figure}

%%%%%%%%%%%%%%%%%%%%%%%%

% use below for double column figures

%%%%%%%%%%%%%%%%%%%%%%%%

\begin{figure}[t] % [h] forces the figure to be output where it is defined in the code (it suppresses floating)
	\centering
	\includegraphics[width=\halfwidth\columnwidth]{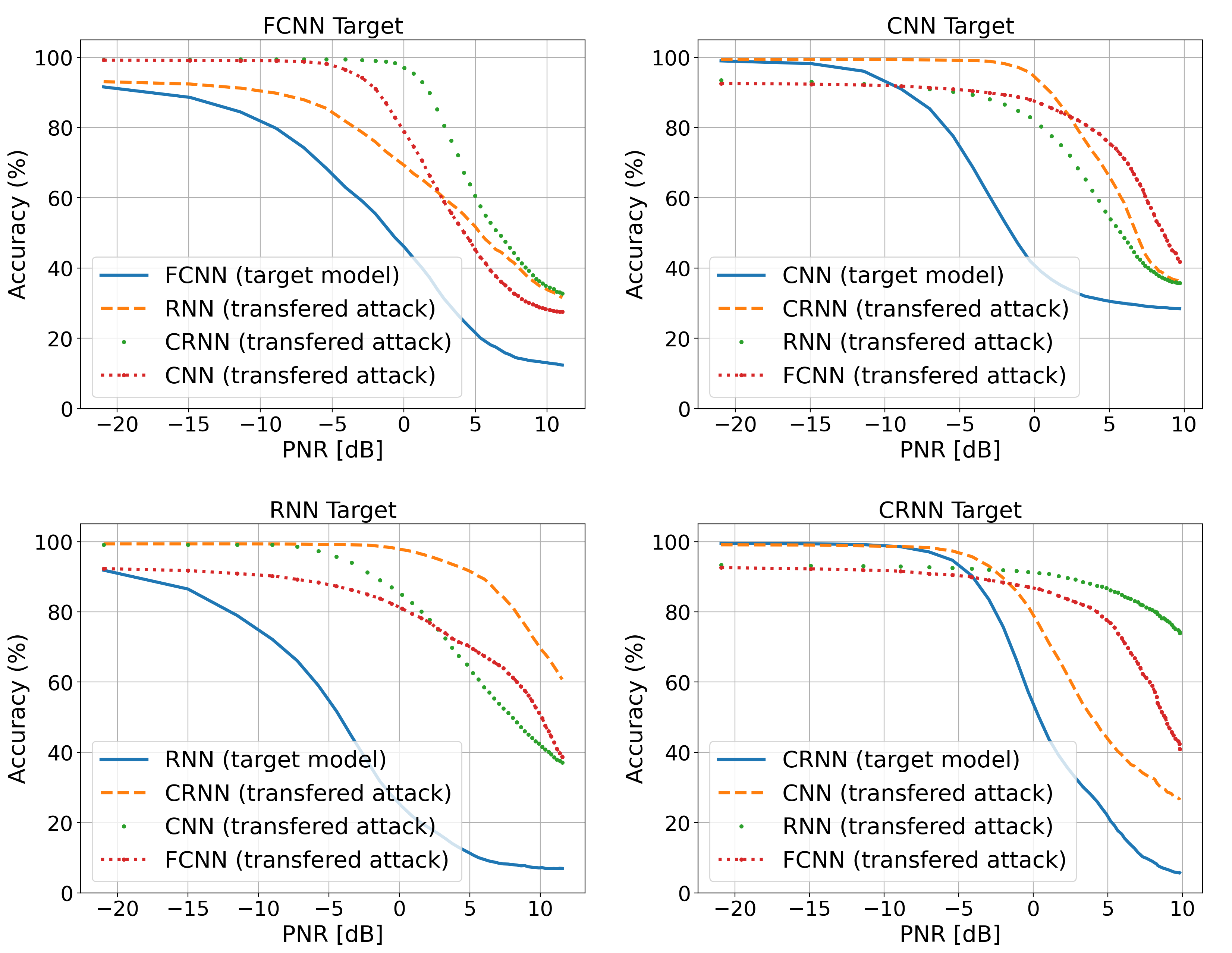}
	\caption{The transferability of the BIM perturbation between classification architectures on Dataset A. We see that the effects of FCNN, CNN, and RNN instantiated attacks are nearly eliminated on the CRNN in the low PNR regime whereas the effects of CRNN instantiated attacks fail to strongly transfer onto RNNs.}
	\label{bim_l2}
\end{figure}

\begin{figure}[t] % [h] forces the figure to be output where it is defined in the code (it suppresses floating)
	\centering
	\includegraphics[width=\halfwidth\columnwidth]{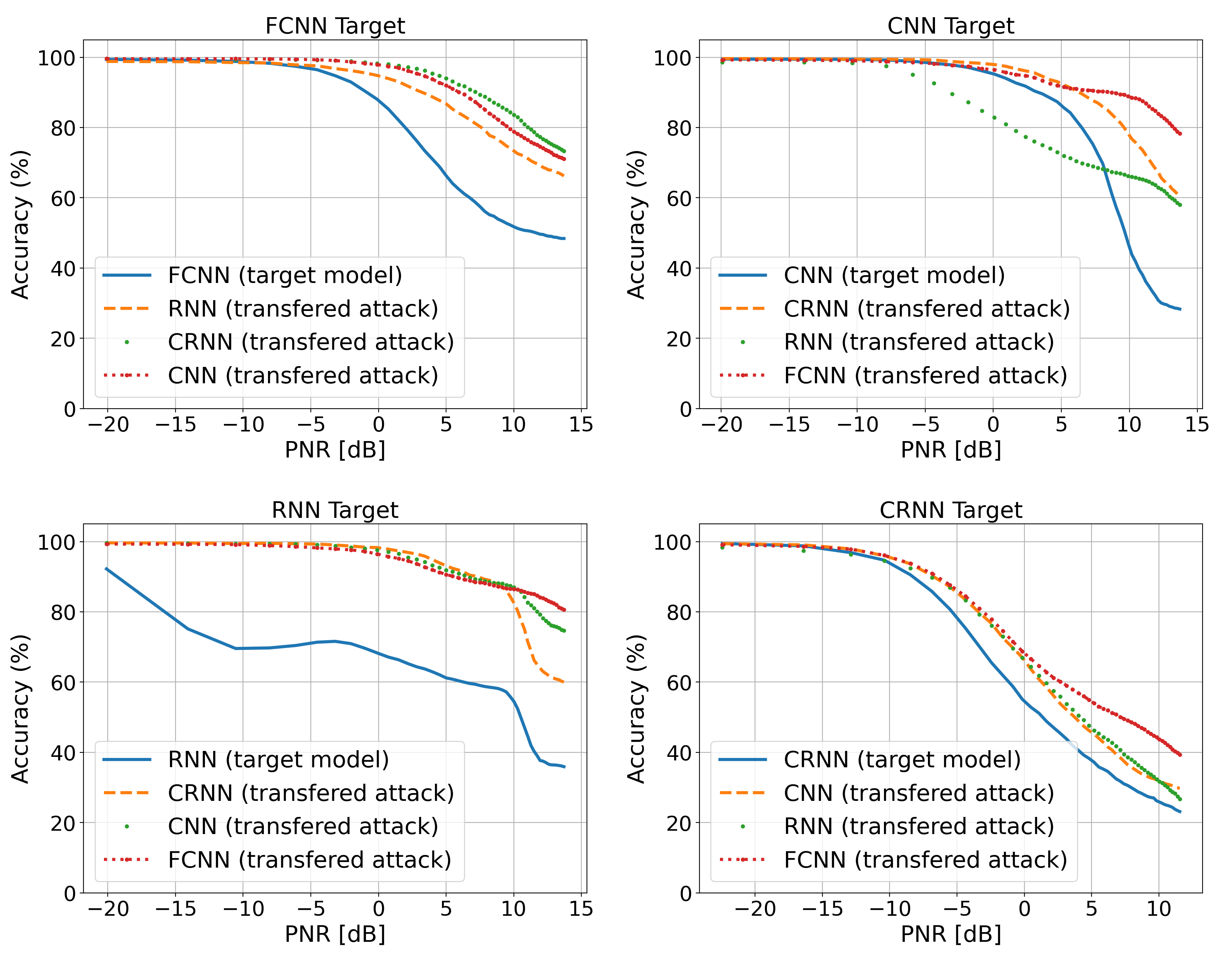}
	\caption{The transferability of the FGSM perturbation between classification architectures on Dataset B. Here, transferrability is smaller than for Dataset A. The FCNN provides the greatest resilience in attacks targeted at the CNN, RNN, and CRNN, whereas the CRNN provides the greatest attack mitigation for interference targeted at the FCNN.}
	\label{fgsm_l2_ds2}
\end{figure}

In Figs. \ref{bim_l2} - \ref{fgsm_l2_ds2}, we demonstrate the ability of our wireless receiver to withstand the effects of transferable adversarial interference in architecture uncertainty environments under the FGSM and BIM attacks for Dataset A and Dataset B. From these figures, we see that the potency of adversarial attacks are mitigated when they are transferred onto architectures differing from the ones used to craft the attacks. In particular, each classifier that experiences a transferred attack has a higher accuracy than the target model for all PNRs. Furthermore, in each case, we find particular classifiers that almost entirely withstand the effects of the additive interference. Specifically, in Fig. \ref{bim_l2}, the CRNN experiences nearly no degradation in the imperceptible PNR region for BIM attacks instantiated on the FCNN, CNN, and RNN, and similarly, the RNN reduces the effects of the CRNN instantiated attacks across nearly the entire considered PNR range. For Dataset B in Fig. \ref{fgsm_l2_ds2}, we observe the same general trends on the FGSM perturbation, except the attacks are less effective overall (due to the lower potency of the FGSM attack compared to the BIM attack), and thus, there is less variation in transferrability between architectures. \emph{This indicates that black box adversarial attacks instantiated on AMC models are not directly transferable between the deep learning classification architectures considered in our wireless receiver}. 

\subsection{Signal Domain Uncertainty Performance}

%%%%%%%%%%%%%%%%%%%%%%%%

% use below for double column figures

%%%%%%%%%%%%%%%%%%%%%%%%

\begin{figure}[t] % [h] forces the figure to be output where it is defined in the code (it suppresses floating)
	\centering
	\includegraphics[width=\halfwidth\columnwidth]{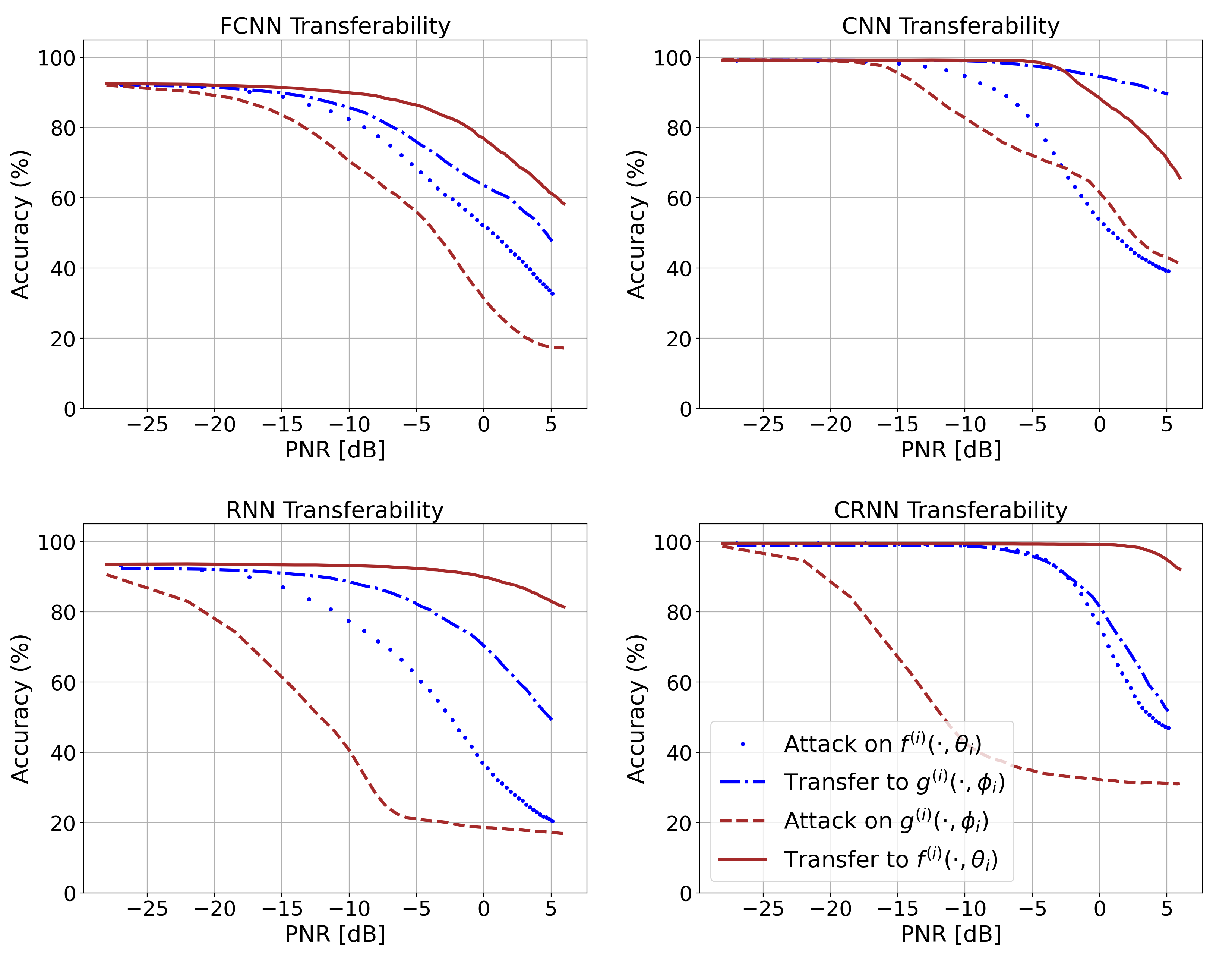}
	\caption{The transferability of the FGSM attack between time and frequency domain classifiers on Dataset A. Both the RNN and CRNN mitigate the effects of the attack targeted at $g^{(i)}(\cdot, \phi_{i})$ to the largest extent when transferred to $f^{(i)}(\cdot, \theta_{i})$, while the frequency domain CNN classifier withstands time domain attacks to the greatest extent.}
	\label{fgsm_dsa}
\end{figure}

\begin{figure}[t] % [h] forces the figure to be output where it is defined in the code (it suppresses floating)
	\centering
	\includegraphics[width=\halfwidth\columnwidth]{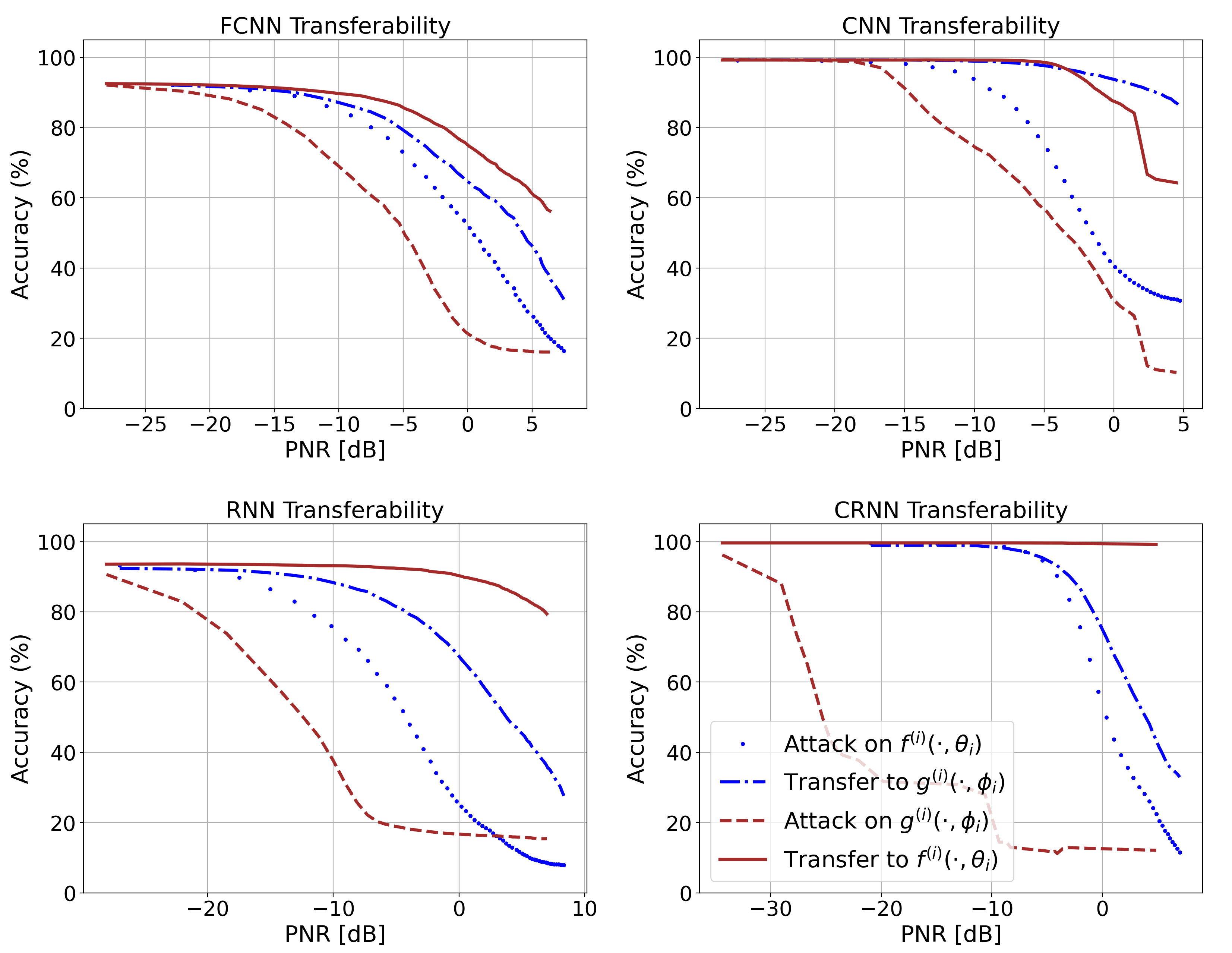}
	\caption{The transferability of the BIM attack between time and frequency domain classifiers on Dataset A. Similar to Fig. \ref{fgsm_dsa}, we see that the RNN and CRNN mitigate frequency domain-based attacks to the greatest extent while the the CNN withstands time domain instantiated attacks to the greatest extent. }
	\label{bim_dsa}
\end{figure}

\begin{table} [t]
\small
\caption{The classification accuracies of the FGSM and BIM attacks between the time and frequency domain classifiers on Dataset B at low  ($-10$ and $-5$ dB), medium  ($0$ dB), and high ($5$ and $10$ dB) perceptibility levels. Consistent with Dataset A, we see that that adversarial perturbations are not directly transferable between signal domains. \label{dsb_trf}}
\centering
\begin{tabular}{| c | c | c | c | c | c |} 
\hline 
\centering 
{\multirow{2}{*}{\makecell{PNR \\ (dB)}}} & {\multirow{2}{*}{Model}} & \multicolumn{2}{c|}{IQ FGSM Attack}  & \multicolumn{2}{c|}{Freq. BIM Attack}  \\
\cline{3-6} 
 & & $f(\cdot, \theta)$ & $g(\cdot, \phi)$ & $g(\cdot, \phi)$  & $f(\cdot, \theta)$ \\
\hline
{\multirow{4}{*}{-10}}  & FCNN & 98.82\% & 98.25\% & 95.20\% & 98.94\% \\
 & CNN & 99.22\% & 99.10\% & 89.38\% & 97.97\%  \\
 & RNN & 69.40\% & 98.90\% & 76.03\% & 84.25\% \\
 & CRNN & 94.71\% & 96.42\% & 99.06\% & 99.47\%  \\
 \hline
{\multirow{4}{*}{-5}}  & FCNN & 96.46\% & 95.97\% & 90.84\% & 98.41\% \\
 & CNN & 98.29\% & 98.41\% & 71.72\% & 93.65\%  \\
 & RNN & 70.01\% & 98.45\% & 74.61\% & 83.16\%  \\
 & CRNN & 75.34\% & 84.25\% & 97.97\% & 99.35\%  \\
 \hline
{\multirow{4}{*}{0}}  & FCNN & 88.03\% & 91.74\% & 69.60\% & 94.79\% \\
 & CNN & 95.20\% & 96.46\% & 62.08\% & 77.87\%  \\
 & RNN & 69.48\% & 95.52\% & 73.35\% & 80.68\%  \\
 & CRNN & 58.74\% & 72.50\% & 88.36\% & 97.84\%  \\
 \hline
{\multirow{4}{*}{5}}  &  FCNN & 66.27\% & 80.84\% & 53.30\% & 79.70\% \\
 & CNN & 85.64\% & 91.66\% & 59.89\% & 64.04\%  \\
 & RNN & 63.67\% & 90.11\% & 70.79\% & 78.44\%  \\
 & CRNN & 37.23\% & 53.09\% & 57.53\% & 90.85\%  \\
\hline
{\multirow{4}{*}{10}}  &  FCNN & 51.74\% & 70.67\% & 43.04\% & 63.63\% \\
 & CNN & 43.81\% & 87.18\% & 61.51\% & 74.17\%  \\
 & RNN & 54.19\% & 85.55\% & 68.02\% & 74.65\%  \\
 & CRNN & 25.87\% & 40.36\% & 15.37\% & 63.38\%  \\

\hline

\end{tabular}
\end{table}

We now evaluate our receiver's ability to withstand transferable adversarial interference between signal domains. Figs. \ref{fgsm_dsa} and \ref{bim_dsa} give the results for the FGSM and BIM attacks on Dataset A. Note that the legend shown in the bottom right subplots of Figs. \ref{fgsm_dsa} and \ref{bim_dsa} apply to all subplots in the figures Each plot shows the transferrability of an attack on the time domain trained classifier, $f^{(i)}(\cdot, \theta_{i})$, to the frequency domain trained classifier, $g^{(i)}(\cdot, \phi_{i})$, and vice versa, for each architecture. In addition, Table \ref{dsb_trf} shows the transferability results on Dataset B for time domain and frequency domain attacks on varying PNRs. 

Fig. \ref{fgsm_dsa}, as well as Table \ref{dsb_trf}, demonstrate the resilience of each trained classifier to withstand FGSM attacks. For both datasets, we see that there are certain architectures for which transferrability from the time domain to the frequency domain, and from the frequency domain to the time domain, are significantly mitigated. In particular, the RNN and CRNN architectures demonstrate the greatest resilience in that they achieve accuracy improvements greater than 70\% and 65\%, respectively, on Dataset A while the attack is less potent on Dataset B for the same DL architectures at 0 dB PNR. Furthermore, the CNN demonstrates significant gains in classification performance when an attack is transferred from the time domain to the frequency domain on both datasets (e.g., improving accuracy from 39.14\% to 89.50\% at 5 dB PNR on Dataset A and from 43.81\% to 87.18\% on Dataset B at 10 dB PNR). The ability of the CNN and CRNN to withstand attacks to the highest degree overall indicates their increased resilience to transferable adversarial interference.

The effectiveness of our proposed defense against the BIM adversarial attack, as shown in Fig. \ref{bim_dsa} and Table \ref{dsb_trf}, consistent with the response of the FGSM attacks: BIM instantiated adversarial attacks are not directly transferable between signal domains. Moreover, the time domain-based CRNN eliminates the degradation effects of the frequency domain instantiated attacks almost completely. However, the degree to which BIM attacks effectively transfer between domains differs between Dataset A and Dataset B, indicating that the mitigation of adversarial attacks in the domain uncertainty environment may be dataset dependent. \emph{Thus, as shown by the instantiation of both considered attacks, our wireless receiver architecture mitigates the transferability of adversarial interference between IQ-based and frequency-based features to a significant degree.} 

%%%%%%%%%%%%%%%%%%%%%%%%

% use below for double column figures

%%%%%%%%%%%%%%%%%%%%%%%%

\begin{figure}[t] % [h] forces the figure to be output where it is defined in the code (it suppresses floating)
	\centering
	\includegraphics[width=\halfwidth\columnwidth]{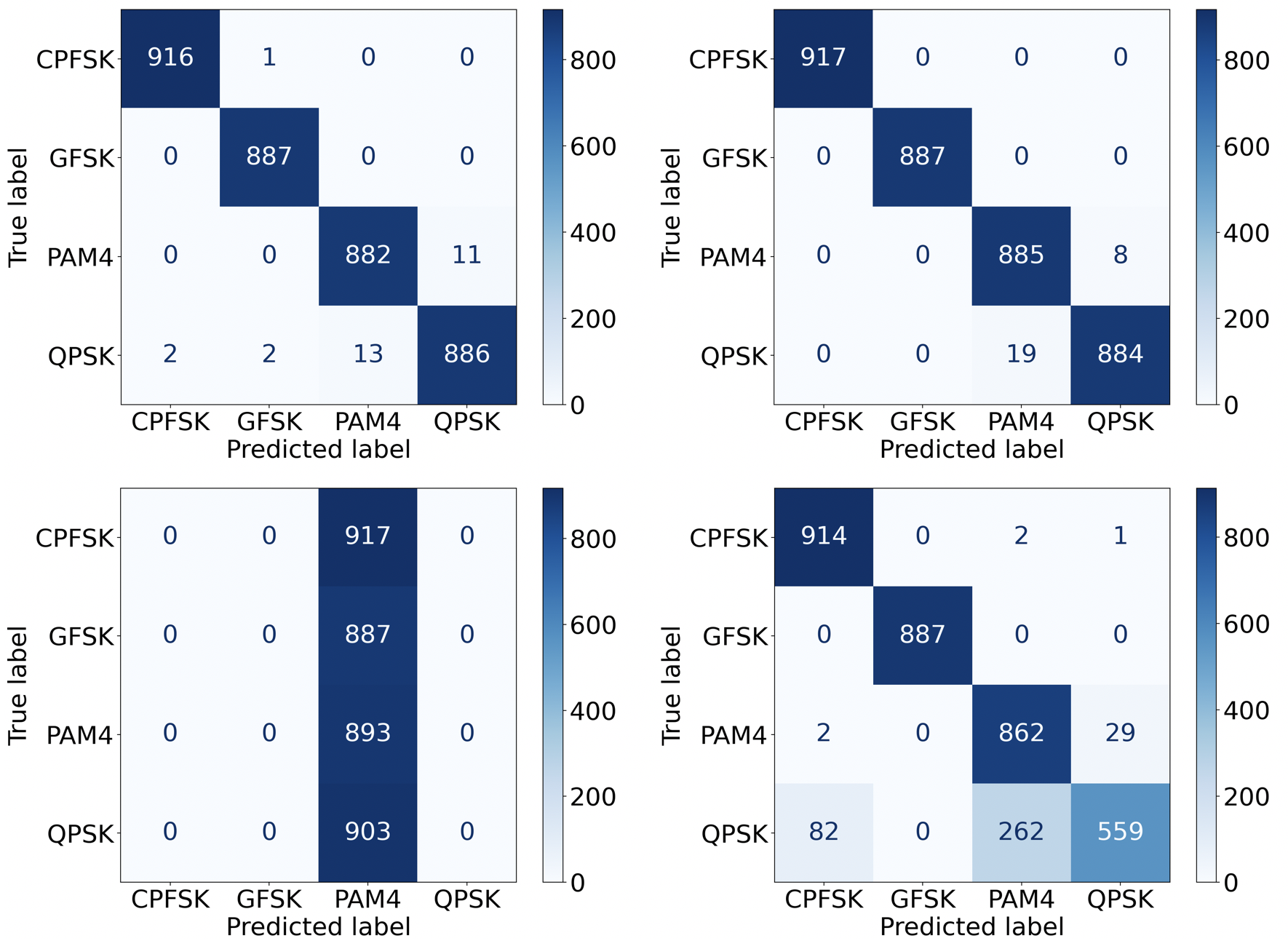}
	\caption{The CNN classifiers on Dataset A, with no interference, achieve roughly equivalent performance using IQ features (top left) and frequency features (top right). Moreover, the instantiation of a 5 dB PNR FGSM attack in the time domain (bottom left) is significantly mitigated on the frequency domain classifier (bottom right), with the largest incongruency being between PAM4 and QPSK.}
	\label{cnn_conf}
\end{figure}

\begin{figure}[t] % [h] forces the figure to be output where it is defined in the code (it suppresses floating)
	\centering
	\includegraphics[width=\halfwidth\columnwidth]{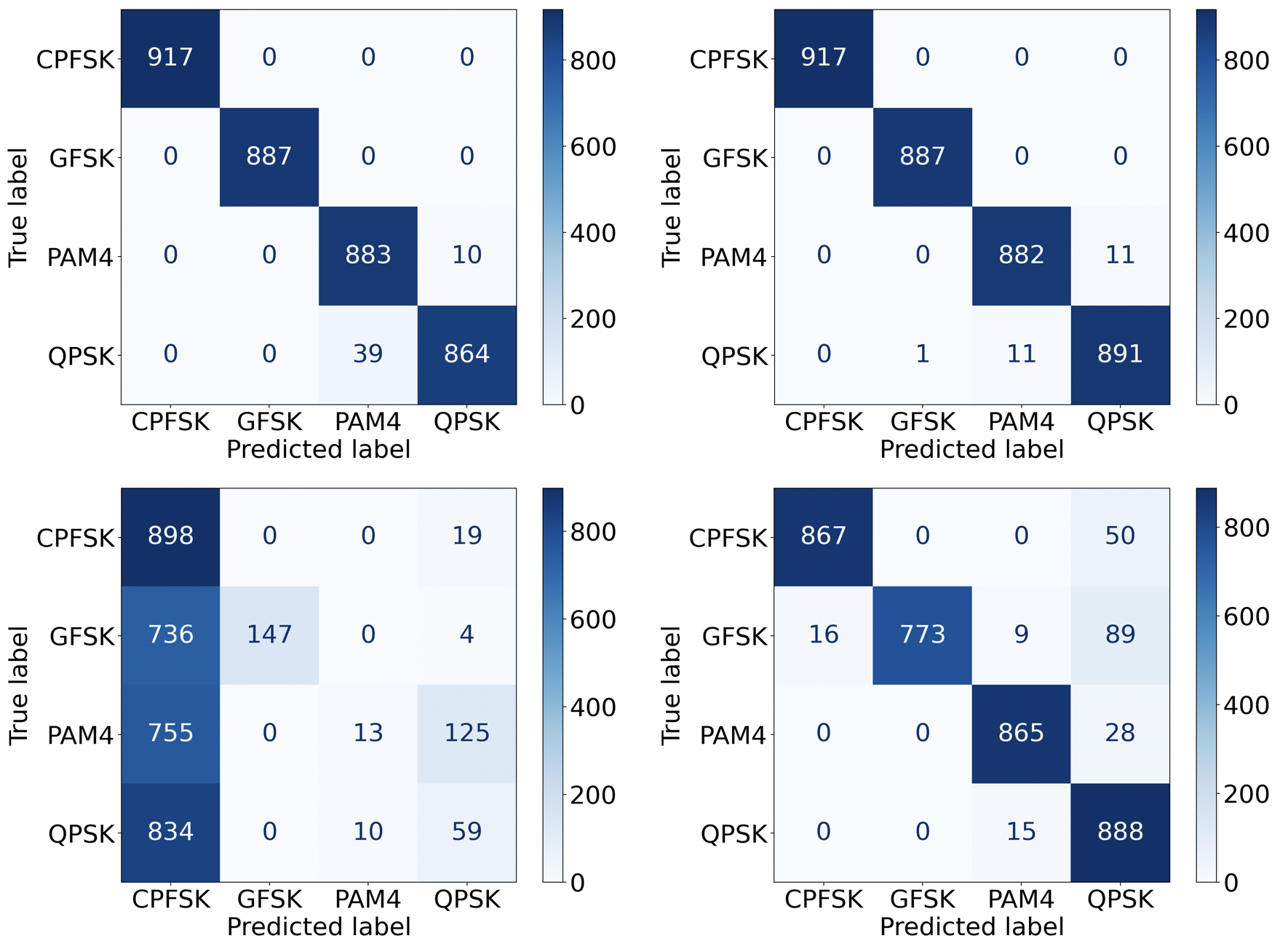}
	\caption{The CRNN classifiers on Dataset A, with no interference, achieve roughly equivalent performance using frequency features (top left) and IQ features (top right), and furthermore, the instantiation of a 5 dB PNR FGSM attack in the frequency domain (bottom left) is significantly mitigated on the time domain feature-based model (bottom right). }
	\label{crnn_conf}
\end{figure}

We analyze the CNN and CRNN's resilience to transferable attacks between signal domains on Dataset A more closely in Figs. \ref{cnn_conf} -- \ref{crnn_conf}. We consider the CNN's performance on a time domain attack compared to the case of no interference in Fig. \ref{cnn_conf}, and the CRNN's performance on a frequency domain attack compared to the case of no interference in Fig. \ref{crnn_conf}.

As shown in Figs. \ref{cnn_conf} and \ref{crnn_conf}, both time and frequency features deliver robust AMC performance in the absence of adversarial interference with classification rates around 99\% for both models. However, at a PNR of 5 dB, the classification rate drops to 39.14\% and 31.03\% when the FGSM attack is transmitted in the time domain, in Fig. \ref{cnn_conf}, and frequency domain, in Fig. \ref{crnn_conf}, respectively. From Figs. \ref{cnn_conf} and \ref{crnn_conf}, we see that the adversarial interference pushes the majority of signals within the classification decision boundaries of the PAM4 modulation constellation for the CNN in the time domain, and the CPFSK constellation for the CRNN in the frequency domain. This is largely due to the nature of the untargeted attack in which the adversary's sole objective is to induce misclassification without targeting a specific misclassified prediction. The attacks are mitigated to a large extent when they are transferred from the time domain to the frequency domain (Fig. \ref{cnn_conf}) and from the frequency domain to the time domain (Fig. \ref{crnn_conf}) with accuracies of 89.50\% and 94.25\%, corresponding to classification accuracy improvements of 50.36\% and 63.22\%, respectively. Frequency domain-based models correctly classify a majority of CPFSK and GFSK modulation schemes corrupted with time domain-based attacks, with the largest incongruency being between PAM4 and QPSK. Time domain-based models, on the other hand, exhibit stronger performance on frequency domain-based attacks, with an overall misclassification rate of 5.75\%. 

\subsection{Assorted Deep Ensemble Defense Performance}

%%%%%%%%%%%%%%%%%%%%%%%%

% use below for double column figures

%%%%%%%%%%%%%%%%%%%%%%%%

\begin{figure}[t] % [h] forces the figure to be output where it is defined in the code (it suppresses floating)
	\centering
	\includegraphics[width=\halfwidth\columnwidth]{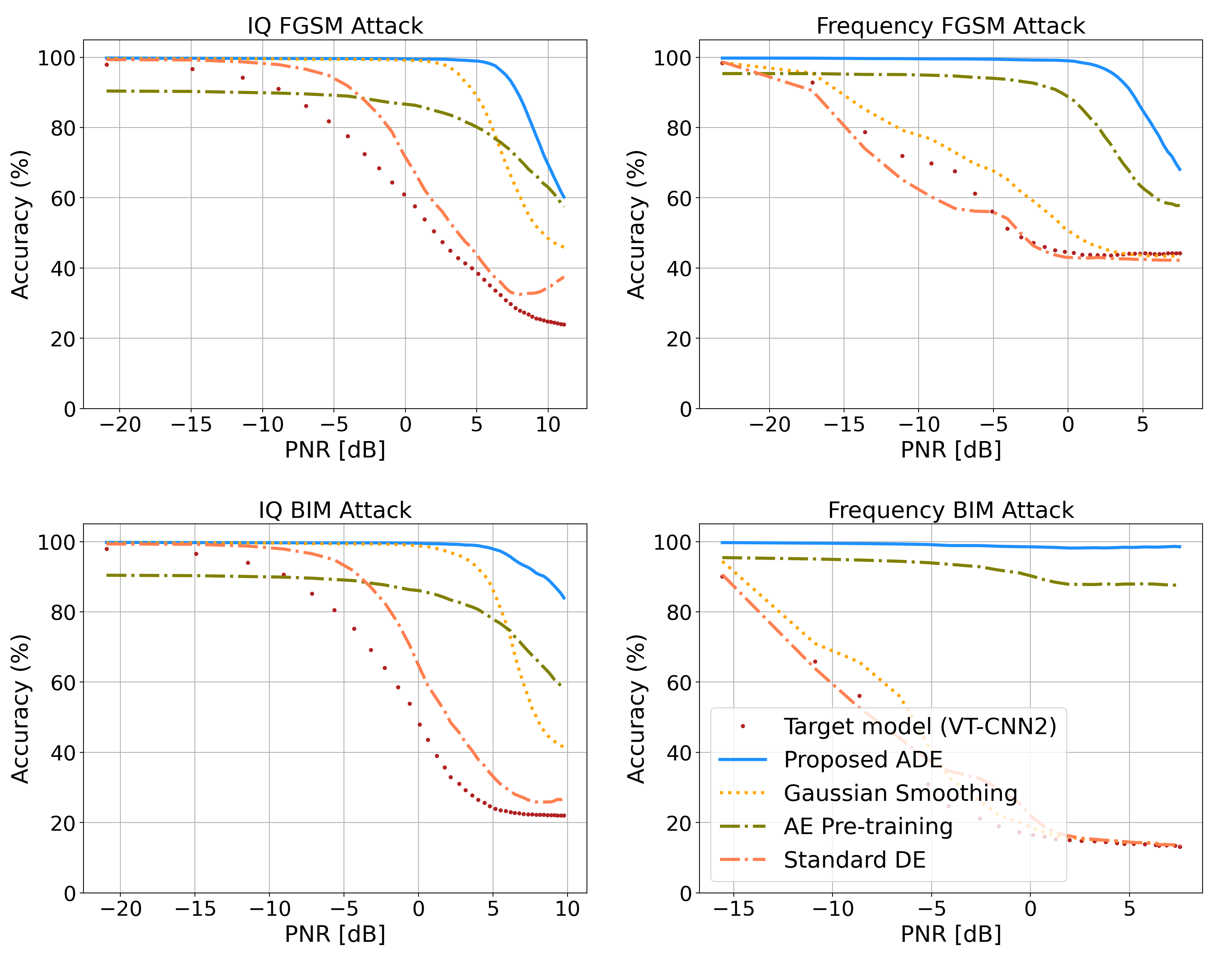}
	\caption{Defense performance of our proposed ADE method compared with three baselines for various types of attack construction on Dataset A. We see that the ADE outperforms the baseline methods on each considered attack for each PNR. }
	\label{defense_dsa}
\end{figure}

\begin{figure}[t] % [h] forces the figure to be output where it is defined in the code (it suppresses floating)
	\centering
	\includegraphics[width=\halfwidth\columnwidth]{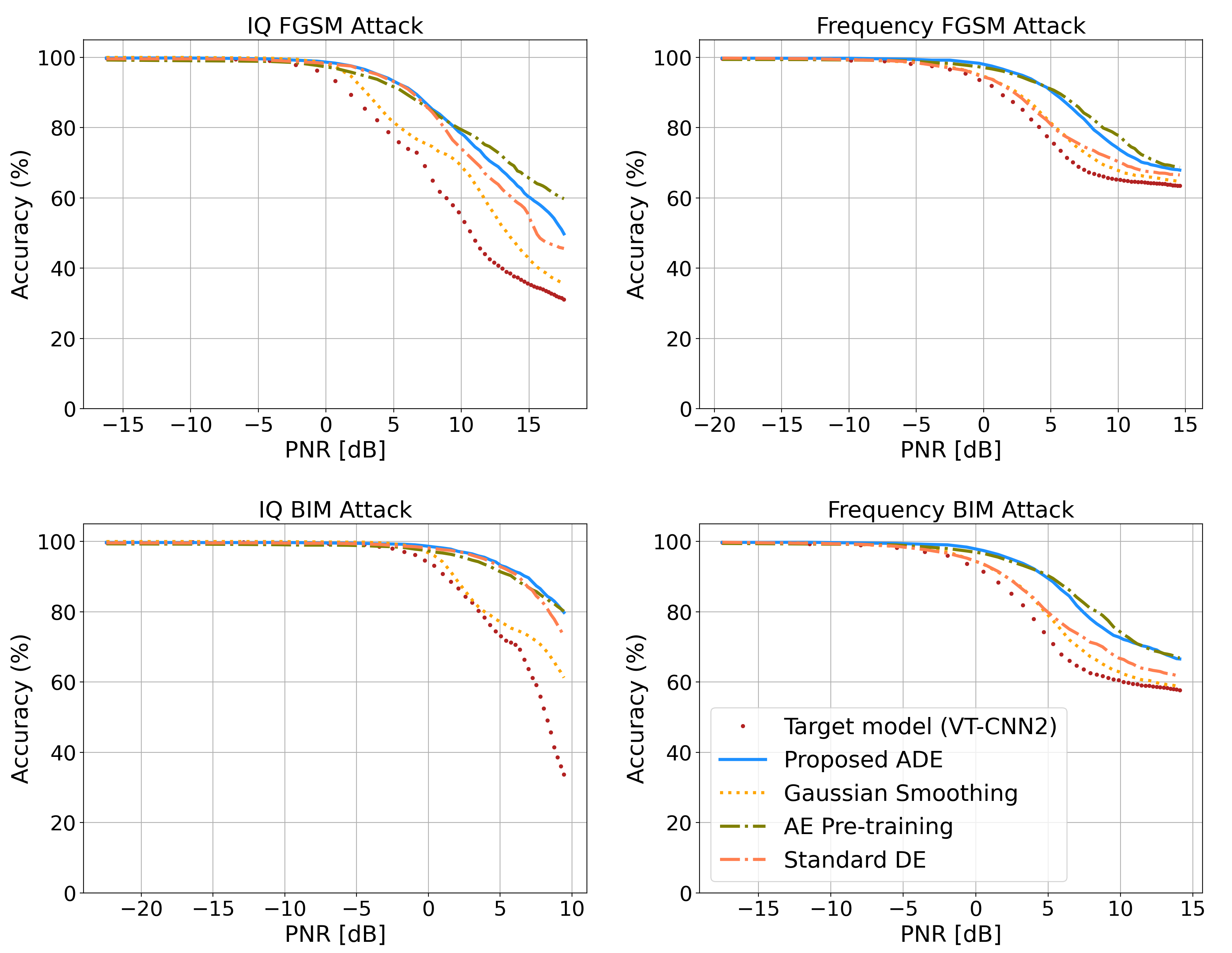}
	\caption{Defense performance of our proposed ADE method compared with three baselines for various types of attack construction on Dataset B. We see that the ADE performs equivalently or outperforms baseline methods on each considered attack.}
	\label{defense_dsb}
\end{figure}

%%%%%%%%%%%%%%%%%%%%%%%%

% use below for single column figures

%%%%%%%%%%%%%%%%%%%%%%%%

% \begin{figure}[!tbp]
%   \centering
%   \begin{minipage}[b]{0.4\textwidth}
%     \includegraphics[width=\textwidth]{defense_dsa_de.png}
%     \caption{Defense performance of our proposed ADE method compared with three baselines for various types of attack construction on Dataset A. We see that the ADE outperforms the baseline methods on each considered attack for each PNR.}
%     \label{defense_dsa}
%   \end{minipage}
%   \hfill
%   \begin{minipage}[b]{0.4\textwidth}
%     \includegraphics[width=\textwidth]{defense_ds2_de.png}
%     \caption{Defense performance of our proposed ADE method compared with three baselines for various types of attack construction on Dataset B. We see that the ADE performs equivalently or outperforms baseline methods on each considered attack.}
%     \label{defense_dsb}
%   \end{minipage}
% \end{figure}

%%%%%%%%%%%%%%%%%%%%%%%%

%%%%%%%%%%%%%%%%%%%%%%%%

% use below for double column figures

%%%%%%%%%%%%%%%%%%%%%%%%

\begin{figure}[t] % [h] forces the figure to be output where it is defined in the code (it suppresses floating)
	\centering
	\includegraphics[width=\halfwidth\columnwidth]{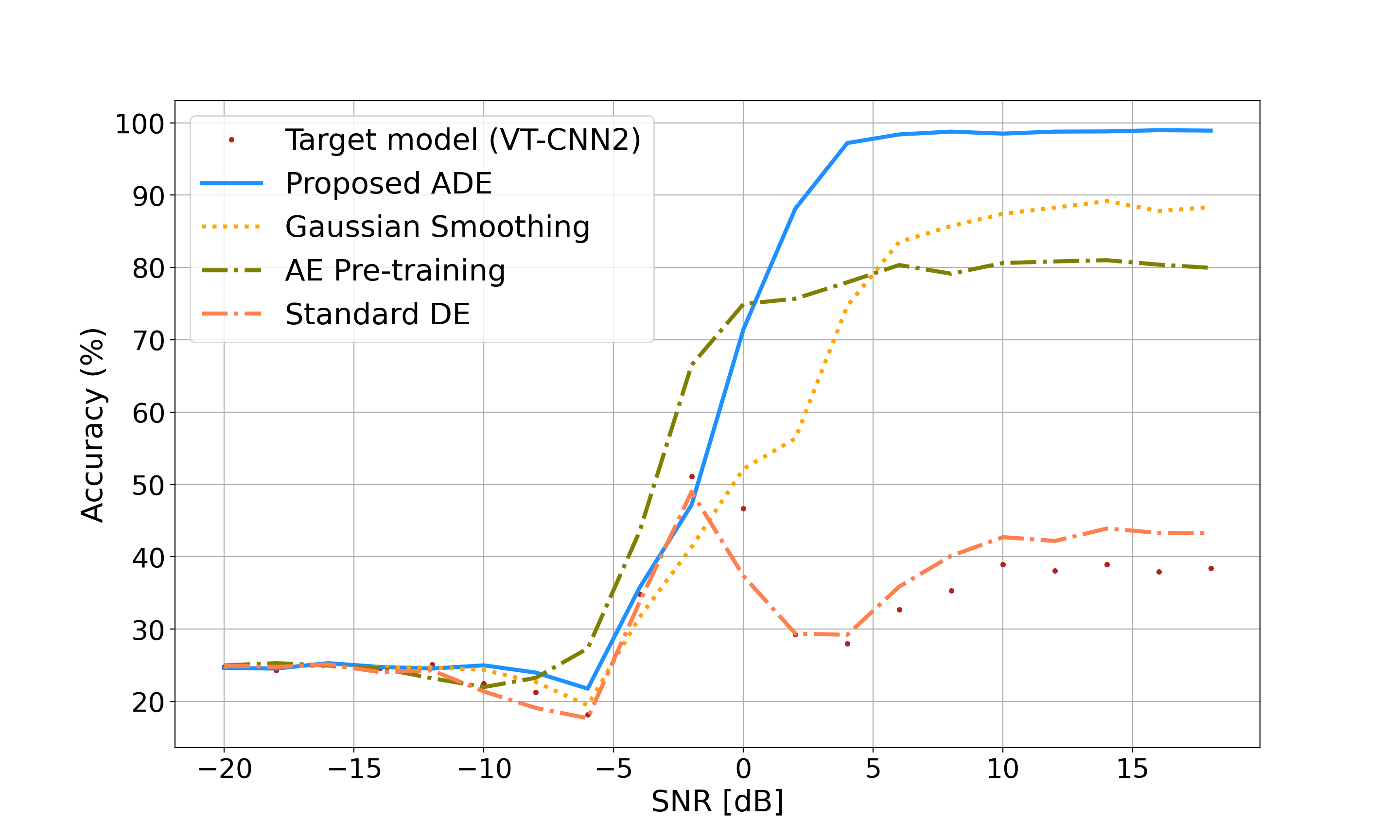}
	\caption{Defense performance of our proposed ADE method compared with each baseline for varying SNRs at a constant PNR of 5 dB on Dataset A. We see that our method outperforms each baseline in the high SNR regime, while each method expectedly performs ineffectively in the low SNR regime. }
	\label{all_snrs}
\end{figure}

%%%%%%%%%%%%%%%%%%%%%%%%

% use below for single column figures

%%%%%%%%%%%%%%%%%%%%%%%%

% \begin{figure}[t] % [h] forces the figure to be output where it is defined in the code (it suppresses floating)
% 	\centering
% 	\includegraphics[width=0.4\columnwidth]{all_snr_5db_pnr.png}
% 	\caption{Defense performance of our proposed ADE method compared with each baseline for varying SNRs at a constant PNR of 5 dB on Dataset A. We see that our method outperforms each baseline in the high SNR regime, while each method expectedly performs ineffectively in the low SNR regime. }
% 	\label{all_snrs}
% \end{figure}

%%%%%%%%%%%%%%%%%%%%%%%%

\begin{table*} [t]
\small
% \caption{The testing accuracy of each considered model on $\mathcal{X}_{te}^{(\cdot)}$ for Dataset A when the CW perturbation is crafted by the adversary with varying confidence levels ($\kappa$) using the VT-CNN2 classifier as the surrogate model. Our proposed ADE mitigates the attack to the greatest extent in comparison to the considered baselines across multiple confidence levels of the attack construction. \label{cw_results}}
\caption{The testing accuracy of each model on $\mathcal{X}_{te}^{(\cdot)}$ for Dataset A when the CW perturbation is crafted by the adversary with varying confidence levels ($\kappa$). Our proposed ADE mitigates the attack to the greatest extent in comparison to the three baselines across multiple confidence levels of the attack. \label{cw_results}}
\centering
\begin{tabular}{| c | c | c | c | c | c | c|} 
\hline 
\centering 
{\multirow{3}{*}{$\kappa$}} & {\multirow{3}{*}{\makecell{Input \\ Features}}} & \multicolumn{5}{c|}{Model Accuracies}  \\
\cline{3-7} 
 & & \makecell{VT-CNN2} & \makecell{\textbf{ADE}} & \makecell{Gaussian \\ Smoothing \cite{gaus_smth}} & \makecell{Autoencoder \\ pre-training \cite{autoencoder_defense}} & \makecell{Standard DE \cite{deep_ensembles}} \\
\hline
{\multirow{2}{*}{0}}  & IQ & 16.53\% & \textbf{99.67\%} & 96.42\% & 87.00\% & 75.25\% \\
 & Frequency & 11.94\% & \textbf{99.58\%} & 40.22\% & 94.02\% & 36.92\% \\
 \hline
{\multirow{2}{*}{10}}  & IQ & 35.02\% & \textbf{98.97\%} & 91.39\% & 83.19\% & 41.94\% \\
 & Frequency & 12.00\% & \textbf{99.19\%} & 13.06\% & 91.75\% & 21.83\% \\
 \hline
{\multirow{2}{*}{20}} & IQ & 60.47\% & \textbf{94.81\%} & 85.17\% & 82.06\% & 61.75\% \\
 & Frequency & 15.67\% & \textbf{98.72\%} & 15.80\% & 90.75\% & 16.78\% \\
  \hline
{\multirow{2}{*}{30}} & IQ & 71.61\% & \textbf{93.31\%} & 79.06\% & 82.08\% & 71.61\% \\
 & Frequency & 16.36\% & \textbf{98.36\%} & 16.58\% & 89.58\% & 16.67\% \\
  \hline
{\multirow{2}{*}{40}} & IQ & 86.14\% & \textbf{96.97\%} & 90.17\% & 86.58\% & 86.31\% \\
 & Frequency & 17.58\% & \textbf{98.50\%} & 17.72\% & 88.33\% & 17.78\% \\
  \hline
{\multirow{2}{*}{50}} & IQ & 94.13\% & \textbf{99.25\%} & 95.89\% & 89.39\% & 94.53\% \\
 & Frequency & 18.44\% & \textbf{98.19\%} & 18.50\% & 87.11\% & 18.47\% \\

\hline

\end{tabular}
\end{table*}

Lastly, we evaluate our proposed assorted deep ensemble (ADE) defense in the overall black box environment, where the adversary crafts an adversarial interference signal using the gradient of a surrogate classifier, as discussed in Sec. III-F. We assume that the adversary uses the VT-CNN2 classifier as the surrogate model, since it is a widely proposed model for DL-based AMC \cite{amc_adv_atk1,adv_amc_detection,dataset,vt_cnn2_3}. The VT-CNN2 classifier is comprised of two convolutional layers containing 256 $1 \times 3$ and 80 $2 \times 3$ feature maps, respectively, followed by a 256-unit  dense layer (as visualized in Fig. \ref{bb_env}). Note that other surrogate models (e.g., the AMC classifier in \cite{ex_clf} or other classifiers contained in our ADE) can be used by the adversary to generate transferable adversarial interference signals, and in such cases, we expect consistent attack mitigation by our ADE due to its demonstrated resilience against transferable adversarial attacks as shown in Sec. IV-C \& IV-D. % Furthermore, the adversary may have partial knowledge about one or more of the models contained in the ADE and, thus use the gradient of that particular classifier to craft the attack. However, since the attack would rely on the transferability property to induce misclassification (since the attack would only be crafted with respect to one classifier), our ADE would mitigate the attack similarly to the complete black box environment considered here due to its resilience to transferable attacks between classification architectures and signal domains as shown in Sec. IV-C \& IV-D. 

To construct our defense strategies outlined in Algorithms \ref{ade_const} and \ref{ade_def}, we use the CNN and CRNN architectures since, as demonstrated from the architecture uncertainty and signal uncertainty environments, they provide the greatest resilience to transferable adversarial interference. We use the following hyper-parameters for constructing the defense for both Dataset A and Dataset B: $M=4$, $k=30$, $\sigma_{\text{IQ}}=0.001$, and $\sigma_{\text{DFT}}=0.005$.

We compare our method to three previously proposed methods for adversarial interference mitigation in AMC: Gaussian smoothing \cite{kim2020channel}, autoencoder pre-training \cite{autoencoder_defense}, and deep ensembles (DE) \cite{deep_ensembles}. Gaussian smoothing consists of retraining a single classifier with samples augmented with random noise in order to improve classification performance on various distortions that may be encountered during deployment such as adversarial examples. Autoencoder pre-training, on the other hand, trains an autoencoder and uses its encoder to calculate a latent space representation of the input data, which is then used to train an AMC classifier, with the rationale being that fewer degrees of freedom (i.e., a lower dimensional representation in the latent space) will prevent misclassifications from adversarial attacks. DEs employ the same classification architecture and use the same data representation for each classifier in a constructed ensemble, with the outputs aggregated to obtain the model's predicted modulation constellation.

Figs. \ref{defense_dsa} and \ref{defense_dsb} show the performance of our ADE when defending the transferred FGSM and BIM attack from the VT-CNN2 classifier on each dataset. On Dataset A, we see that our proposed defense outperforms the baselines for all attacks and PNRs considered, with almost no degradation in classification for PNRs below 2 dB. For example, our ADE method improves the classification accuracy from 22.23\% to 90.53\% on the time domain BIM attack at 8 dB PNR, whereas Gaussian smoothing, autoencoder pre-training, and standard DEs achieve classification accuracies of 47.86\%, 65.33\%, and 25.94\%, respectively, on the same attack. In comparison to DEs, we find that incorporating diversity into the classifiers comprising a deep ensemble, as we propose in our ADE, is necessary for mitigating transferable adversarial attacks. In our ADE, this diversity stems from the gradient mismatch between the adversary's classifier used to craft the attack and the classifiers used in the ADE at the receiver. In this regard, the gradient mismatch results in an attack that is less potent on the ADE, thus resulting in higher classification rates on adversarially perturbed inputs. In addition, for lower PNRs, we see that Gaussian smoothing outperforms autoencoder pre-training for time domain attacks, while autoencoder pre-training is significantly better than Gaussian smoothing for defending frequency domain instantiated attacks. 

For Dataset B, we see that the performances of each defense are generally closer, which is consistent with our prior observations that attacks are less potent on this dataset. The performance of our ADE is comparable to autoencoder pre-training, while Gaussian smoothing performs similarly to no defense mitigation. Finally, we see that our proposed defense continues to mitigate the effects of attacks even after the received signal is masked by the perturbation ($\text{PNR} > 0$ dB).  %Finally, we see that in each case, on both datasets, the classification accuracy degrades at high PNR due to the received signal being masked by the perturbation. 

In addition, Table \ref{cw_results} shows the performance of our ADE defense against the CW attack crafted on the VT-CNN2 classifier under varying confidence levels $\kappa$. Overall, we see that our proposed ADE defense significantly mitigates the effects of the CW perturbations across multiple confidence levels in a complete black box environment. By contrast, each of the baselines exhibits significant performance degradation as $\kappa$ and/or the input features are varied. Similar results were observed on Dataset B (omitted for brevity). 

We now consider the effect of varying the SNR while holding the PNR constant. Fig. \ref{all_snrs} shows the classification performance of our method, as well as each considered baseline, for the SNR range of $[-20, 18]$ dB while holding the PNR constant at 5 dB. The results are shown here for the FGSM attack on the Dataset A (we find them to be consistent for other cases as well). We see that, in comparison to the SNR of 18 dB at which the results have been presented in this paper, the performance obtained by our method is consistent when the SNR is greater than 2 dB. At SNRs lower than 2 dB, classification performance is overall degraded due to poor signal quality (consistent with prior work, e.g., \cite{amc_dl1,amc_dl2,amc_dl3}). In these cases, the classification performance, even of the improved gains in accuracy, are not very effective due to the poor signal environment induced from low SNR waveforms. Future work must consider methods to improve AMC performance in the low SNR regime before evaluating the effects of adversarial attacks in this setting.

\section{Conclusion}

Deep learning (DL) has recently been proposed as a robust method to perform automatic modulation classification (AMC). Yet, deep learning AMC models are vulnerable to adversarial attacks, which can alter a trained model's predicted modulation constellation with relatively little input power. Furthermore, such attacks are transferable, which allows the interference to degrade the performance of several classifiers simultaneously. In this work, we developed a novel wireless transmission receiver architecture -- consisting of both time and frequency domain feature-based classification models -- which is capable of mitigating the transferability of adversarial interference in black box environments. Specifically, we showed that our models are resilient to transferable adversarial attacks between DL classification architectures and between the time and frequency domain, where convolutional neural networks (CNNs) and convolutional recurrent neural networks (CRNNs) demonstrated the greatest degree of mitigation. Using these insights, we proposed our assorted deep ensemble defense, which defends a wireless receiver from complete black box adversarial perturbations. We found that our proposed method is capable of mitigating adversarial AMC attacks to a greater extent than previously proposed methods, thus increasing the robustness of deep learning AMC receivers from malicious behavior. 

%Specifically, we showed that our proposed frequency feature-based deep learning classifiers are resilient to transferable adversarial interference instantiated on traditional time domain in-phase and quadrature (IQ) feature-based models and vice versa. We also demonstrated the resilience of domain-specific classification models to withstand the effects of adversarial attacks crafted on a differing classification architecture, with the convolutional neural network (CNN) and convolutional recurrent neural network (CRNN) achieving the highest degrees of resilience in different cases. Finally, we proposed our assorted deep ensemble defense, which defends a wireless receiver from black box perturbations that are crafted using imperceptible magnitudes of additive interference. We found that our proposed method is capable of mitigating black box interference to a greater extent than previously proposed methods, thus increasing the robustness of wireless AMC channels from malicious behavior. 

% future work: availability of more datasets 

% time domain attacks are not transferable to frequency based features and vice versa

\bibliography{references}

\bibliographystyle{IEEEtran}

% \end{IEEEbiography}


% % insert where needed to balance the two columns on the last page with
% %\newpage


\end{document}